\documentclass[twocolumn,superscriptaddress,showpacs, longbibliography, aps, prx]{revtex4-2}
\usepackage{dcolumn}
\usepackage{physics}
\usepackage{bm}
\usepackage{amsmath,amssymb}
\usepackage{mathtools}
\usepackage[normalem]{ulem} 
\usepackage{xcolor}
\usepackage{hyperref}
\hypersetup{colorlinks=true,breaklinks,linkcolor=blue,urlcolor=blue,citecolor=blue}
\def\vec#1{\boldsymbol #1}

\begin{document}

\title{Thermal Hall transport in Kitaev spin liquids}
\author{Tsuyoshi Okubo}
\email{t-okubo@phys.s.u-tokyo.ac.jp}
\affiliation{Institute for Physics of Intelligence, University of Tokyo, Tokyo, 113-0033, Japan}
\author{Joji Nasu}
\affiliation{Department of Physics, Tohoku University, Sendai 980-8578, Japan}
\author{Takahiro Misawa}
\affiliation{Beijing Academy of Quantum Information Sciences, Haidian District, Beijing 100193, China}
\affiliation{Institute for Solid State Physics, University of Tokyo, 5-1-5 Kashiwanoha, Kashiwa, Chiba 277-8581, Japan}
\author{Yukitoshi Motome}
\affiliation{Department of Applied Physics, University of Tokyo, Tokyo, 113-8656, Japan}

\begin{abstract}
The thermal Hall conductivity---a thermal current analog of the Hall conductivity for electric current---is a useful detector for exotic quasiparticles emerging from quantum many-body effects, such as Majorana fermions.
In particular, the half-integer quantization of the thermal Hall conductivity at the zero temperature limit has been regarded as direct evidence of Majorana fermions in their topological Chern insulating state. 
However, it remains unclear how the thermal Hall conductivity behaves at finite temperatures, even in the celebrated Kitaev model on a honeycomb lattice under an external magnetic field, which is one of the fundamental models for realizing a Majorana Chern insulator in quantum magnets, despite its crucial importance in reconciling conflicting experimental observations. 
Here, we investigate the thermal Hall conductivity in the Kitaev model with additional interactions under a magnetic field, employing a finite-temperature tensor network method benchmarked by a thermal pure quantum state technique. 
We find that the thermal Hall conductivity divided by temperature, $\kappa_{xy}/T$, significantly overshoots the value of the half-integer quantization and exhibits a pronounced hump while decreasing temperature. 
Moreover, we show that the field-direction dependence of $\kappa_{xy}/T$ is consistent with the sign of the Chern number associated with the Majorana fermions across a wide range of magnetic fields. 
We also demonstrate that the additional off-diagonal interactions, known as the $\Gamma$ and $\Gamma^{\prime}$ terms, considerably affect $\kappa_{xy}/T$. 
In particular, we show that positive $\Gamma$ and negative $\Gamma^{\prime}$ lead to a remarkable enhancement in the intermediate temperature region. 
From the comparison with the classical counterpart, we reveal that the effects of the $\Gamma$ term go beyond the classical picture, indicating significant quantum fluctuation effects, while those of the $\Gamma^\prime$ term are well captured at the classical level. 
These comprehensive analyses indicate that the enhanced thermal Hall response is consistently explained by dominant contributions from topological Majorana fermions, even within the polarized regime beyond the critical field. 
Our approach not only establishes a robust theoretical framework for understanding the thermal Hall transport in Kitaev materials such as $\alpha$-RuCl$_{3}$, but also offers a promising pathway to bridge the gap between theories and experiments across a wide range of strongly correlated materials.
\end{abstract}
\maketitle
\section{Introduction}\label{sec:introduction}
Quantum many-body systems are complex systems in which interactions between a large number of particles lead to emergent phenomena that cannot be understood by considering individual particles alone. 
One fascinating aspect of these systems is the emergence of fractional excitations.
Strongly correlated particles can produce elementary excitations carrying quantum numbers that are fractions of those of the underlying constituent particles.  
For instance, strongly correlated electrons confined in a two-dimensional space under a strong magnetic field exhibit exotic excitations with fractions of elementary charge, known as the fractional quantum Hall effect~\cite{Laughlin1983, Wen1995}. 
These fractional excitations are significant because they reveal deeper insights into the nature of quantum entanglement and topological order in quantum many-body systems~\cite{Wen2007Book,Wen1990, KitaevP2006, LevinW2006}. 
They also have potential applications in quantum computing, particularly in the development of topological quantum computers that leverage the robustness of these fractional states and their peculiar statistics, as discussed for non-Abelian anyons in the fractional quantum Hall effect.

A paradigmatic example of fractionalization in magnets is found in quantum spin liquids (QSLs). 
QSLs are enigmatic states of insulating quantum spin systems, which do not exhibit any spontaneous symmetry breaking even at zero temperature~\cite{Anderson1973, Wen1991, Kitaev2006, Balents2010}. 
This intriguing possibility was first proposed by Anderson for a resonating valence bond (RVB) state, which is given by a superposition of spin-singlet coverings of the entire lattice~\cite{Anderson1973}.
In this RVB state, spin excitations are fractionalized into two types of emergent quasiparticles: spinons, which are particlelike excitations carrying spin $S=1/2$ and no charge, and visons, which are topological excitations carrying no spin and no charge in the traditional sense~\cite{Wen2007Book, SenthilF2000}.
This proposal has stimulated a deeper understanding of quantum many-body physics, including the concepts of emergent gauge fields and the projective symmetry group~\cite{SenthilF2000,Wen2002,Wen2003}. 
Different types of QSLs can produce different sets of emergent fractional excitations, and the search for such QSLs has been one of the central issues of modern condensed matter physics. 

Among various types of QSLs, the Kitaev QSL has attracted significant interest since its proposal by Kitaev~\cite{Kitaev2006}. 
The Kitaev QSL is realized in the ground state of the Kitaev model on a honeycomb lattice, which features frustrated bond-dependent anisotropic interactions. 
The model is exactly solvable and exhibits fractional quasiparticles in its low-energy excitations. 
Two types of emergent quasiparticles appear: itinerant Majorana fermions and localized $Z_2$ fluxes, corresponding to spinons and visons in Anderson's RVB state, respectively.
Since it was shown that the model can be realized in Mott insulating systems under strong spin-orbit coupling~\cite{Jackeli_PRL2009}, the Kitaev QSL has been extensively studied both theoretically and experimentally~\cite{WinterTDBSGV2017, TakagiTJKN2019, KnolleM2019, JanssenV2019, Motome_JPSJ2020, TrebstH2022, RousochatzakisPLK2024}. 
Despite several indications in candidate materials, detecting the fractional excitations in the Kitaev QSL remains a challenge because the emergent quasiparticles carry neither charge nor spin, and do not show any characteristic behaviors in response to electric or magnetic fields.  

The thermal Hall measurement is a promising tool for detecting the emergent Majorana fermions, as these itinerant quasiparticles carry heat~\cite{Kitaev2006}. 
In the Kitaev model, a perturbatively-introduced magnetic field opens a gap at the Dirac-like nodal points in the Majorana band structure, and makes the system a topologically-nontrivial Majorana Chern insulator. 
Accordingly, a chiral edge current appears, protected by the band topology characterized by a nonzero Chern number $\nu$. 
Because the chiral edge current is an energy current of Majorana fermions, the thermal Hall conductivity $\kappa_{xy}$ divided by temperature $T$ is quantized in the limit of $T\to 0$ as
\begin{align}
{\frac{\kappa_{xy}}{T} = \frac{1}{2} \frac{\pi k_{\mathrm{B}}^2}{6\hbar}}\nu,
\label{eq:k_xy}
\end{align}
where $k_{\mathrm{B}}$ and $\hbar$ denote the Boltzmann and Dirac constants, respectively. 
Here, $\nu$ takes $\pm 1$ depending on the direction of the magnetic field. 
The coefficient of $1/2$ is the consequence of the fractionalization; each Majorana fermion carries half the degrees of a conventional Dirac fermion. 
Thus, this quantized value is half of that for electrons with the same band topology, and hence, it is called the half-integer quantization. 
Detecting this peculiar behavior offers direct evidence of the Majorana fermions in the Kitaev QSL.

Strikingly, such a half-integer quantization was reported in the thermal Hall experiment for a quasi-two-dimensional van der Waals magnet $\alpha$-$\mathrm{RuCl_3}$ under a magnetic field~\cite{Kasahara_Nature2018}. 
Although $\alpha$-$\mathrm{RuCl_3}$ shows magnetic zigzag order at zero magnetic field~{\cite{Sears_PRB2015,Williams_PRB2015,Cao_PRB2016}}, inelastic neutron scattering measurements suggest that applying a magnetic field drives the system into a QSL~\cite{Banerjee_npjQ2018}. 
The half-integer quantization observed in this field regime suggests that topologically-nontrivial Majorana bands are realized in this material. 
Similar behavior was observed even when subjected to an in-plane magnetic field, in contrast to the conventional thermal Hall effect of electrons that requires an out-of-plane field~\cite{Yamashita_PRB2020,Bruin_NPhys2022}. 
It was also shown that field-direction dependence of the thermal Hall conductivity is consistent with that in the Kitaev model, suggesting that the Majorana fermion bands in the real material have the same topology predicted by the Kitaev model~\cite{Yokoi_Science2021}.

However, there remains controversy in interpreting the observed thermal Hall transport as evidence of the Kitaev QSL. 
For example, the observed thermal Hall conductivity exhibits an overshooting behavior while decreasing temperature before converging to the half-integer quantized value. 
The origin of this behavior remains unresolved, although contributions from phonons~\cite{Ye2018Quantization,Vinkler2018} and visons~\cite{Joy2022} were proposed. 
In addition, real compounds, including $\alpha$-$\mathrm{RuCl_3}$, may contain additional interactions not included in the Kitaev model, such as the Heisenberg interaction~\cite{Chaloupka_PRL2010,Chaloupka_PRL2013}, and the symmetric off-diagonal $\Gamma$ and $\Gamma'$ interactions~\cite{Rau2014,Rau2014pre,Winter2016}, whose effects remain elusive. 

Furthermore, several subsequent experiments on $\alpha$-$\mathrm{RuCl_3}$ have reported the absence of half-integer quantization. 
For instance, instead of the quantization in the transverse component, quantum oscillations were observed in the longitudinal thermal conductivity~\cite{Czajka_NPhy2021}. 
Additionally, other origins of the thermal Hall response were discussed, e.g., topological magnons~\cite{Czajka_NMat2023,ChernZK2021,ZhangCK2021, McClartyDGRPMP2018,Joshi2018}, para-magnon~\cite{Hentrich_PRB2019}, and phonons~\cite{Lefra_PRX2022, OhN2025}. 
Recent theoretical and experimental studies have also discussed geometry dependence of the thermal Hall conductivity~\cite{Halasz2025,ZhangHGJWTMY2025}.
To elucidate the origin of the thermal Hall conductivity observed in experiments and to reconcile the conflicting experimental findings, it is crucial to clarify the contributions to its finite-temperature behavior from the Kitaev model and its extensions.

For the Kitaev model without other additional interactions, efficient numerical simulation methods have been developed for studying thermodynamic properties and spin dynamics at finite temperatures, utilizing the Majorana fermion representation~\cite{NasuUM2014,NasuUM2015,YoshitakeNM2016,YoshitakeNKM2017,YoshitakeNM2017}. 
This approach explicitly leverages the emergent quasiparticle picture in fractionalization. 
Although these methods are limited to zero magnetic field due to the negative sign problem, a Majorana-based quantum Monte Carlo simulation has been performed for thermal transport in an effective model derived by a perturbation expansion with respect to the magnetic field, which avoids the sign problem by remaining within the Majorana fermion framework~\cite{NasuYM2017}. 
This study revealed a non-monotonic temperature dependence in $\kappa_{xy}/T$, featuring a weak peak at high temperature, but the result cannot explain the overall temperature dependence observed in experiments, including the overshooting behavior at low temperatures. 
For the regime beyond the perturbation, a continuous-time quantum Monte Carlo simulation in the original spin basis has been developed~\cite{YoshitakeNKM2020}, but it also suffers from the negative sign problem at low temperatures and was not applied to the study of the thermal Hall transport. 
In addition, because the standard Kubo formula based on the linear-response theory requires information on dynamical physical quantities, reliable numerical calculations of thermal transport for microscopic models are challenging.

The effects of non-Kitaev interactions beyond the pure Kitaev model have also been extensively investigated. 
For instance, perturbation expansions demonstrated that the $\Gamma'$ 
interaction introduces an additional contribution to the gap of Majorana fermion band~\cite{TakikawaF2020}. 
This gap increases (decreases) with a negative (positive) $\Gamma'$ interaction, potentially influencing the temperature dependence of $\kappa_{xy}/T$. 
In contrast, the effect of the $\Gamma$ interaction appears at higher-order expansions~\cite{YamadaF2021}, suggesting a weaker influence on the thermal Hall transport compared to that of $\Gamma'$. 
A similar perturbative analysis predicted quantum phase transitions, leading to nontrivial temperature dependence of $\kappa_{xy}/T$~\cite{GoJM2019}. 
However, all these analyses are based on perturbation theory, which is valid only for small $\Gamma$ and $\Gamma'$ in the zero temperature limit and do not take into account the contributions from $Z_2$ flux excitations. 
Although the actual values of $\Gamma$ and $\Gamma'$ vary among candidate materials, they are not sufficiently small to justify the use of a perturbative approach; for example, $\Gamma$ is considered to be comparable to the dominant Kitaev interaction in $\alpha$-RuCl$_3$~\cite{Suzuki2018,laurell2020dynamical,Maksimov2020}. 
Recently, an unbiased calculation using a tensor network method was conducted to study the thermal Hall conductivity in the Kitaev-Heisenberg model~\cite{KumarT2023}, but the definition of energy current remains incomplete, as detailed later. 
Thus, to achieve a comprehensive understanding of the effects of the non-Kitaev interactions on the thermal Hall transport, it is essential to explore a theoretical framework that includes a proper definition of energy current and goes beyond perturbation approaches.

In this paper, we investigate the thermal Hall transport for an extended Kitaev model in a magnetic field, which includes the symmetric off-diagonal interactions, $\Gamma$ and $\Gamma^\prime$, using a finite-temperature tensor network method benchmarked with a thermal pure quantum state method. 
For comparison, we also perform classical Monte Carlo simulations in the classical limit of the model. 
We properly define the energy current from the model Hamiltonian and calculate the edge current in the systems with open boundaries. 
This treatment does not assume a specific origin for the thermal Hall transport, and includes all contributions in this spin model on an equal footing. 
Furthermore, this approach allows us to evaluate the thermal Hall conductivity without the explicit computation of dynamical quantities.
We find that $\kappa_{xy}/T$ calculated for the pure Kitaev model shows a clear overshooting behavior, forming a hump at an intermediate temperature, which cannot be obtained by perturbation theory. 
We also investigate the effects of $\Gamma$ and $\Gamma^\prime$, showing that these interactions considerably affect the thermal Hall conductivity in both its magnitude and sign.
In particular, we find that the thermal Hall conductivity is significantly enhanced by positive $\Gamma$ and negative $\Gamma^\prime$ in an intermediate temperature range. 
Furthermore, by comparing with the results of classical Monte Carlo simulations, we reveal the strong quantum nature of the $\Gamma$ contributions. 
Our comprehensive analyses suggest that these nonperturbative behaviors of the thermal Hall conductivity primarily stem from topological Majorana fermions. 
This insight would contribute to a deeper understanding of the experimental observations and offer a potential resolution to the existing discrepancies. 

The organization of this paper is as follows. 
In Sec.~\ref{sec:model}, we introduce the extended Kitaev model and give a brief overview of previous studies on this model. 
In Sec.~\ref{sec:method}, we explain the procedure for calculating the thermal Hall conductivity in the extended Kitaev model. 
We also introduce the numerical methods used in this study, i.e., the exponential tensor renormalization (XTRG) method, the canonical thermal pure quantum (cTPQ) state method, and classical Monte Carlo simulation.
In Sec.~\ref{sec:pureKitaev}, we present numerical results on the temperature dependence of the thermal Hall conductivity, along with other thermodynamic quantities, for the pure Kitaev model under a magnetic field along the [111] direction. 
We also examine the dependence on the field direction and compare the results with predictions from perturbation theory. 
In Sec.~\ref{sec:Gamma}, we explore the effects of the off-diagonal interactions $\Gamma$ and $\Gamma^{\prime}$. 
In Sec.~\ref{sec:summary_of_temp_field}, we summarize the temperature and field dependence across a wide range of $\Gamma$ and $\Gamma'$ interactions.
In Sec.~\ref{sec:classicalMC}, we analyze the classical limit of the Kitaev model and discuss the significance of quantum contributions by comparing the classical and quantum results.
Section~\ref{sec:Summary} is devoted to a summary.

\section{Model}
\label{sec:model}
\begin{figure}[tbh]
  \begin{center}
    \includegraphics[width=\linewidth]{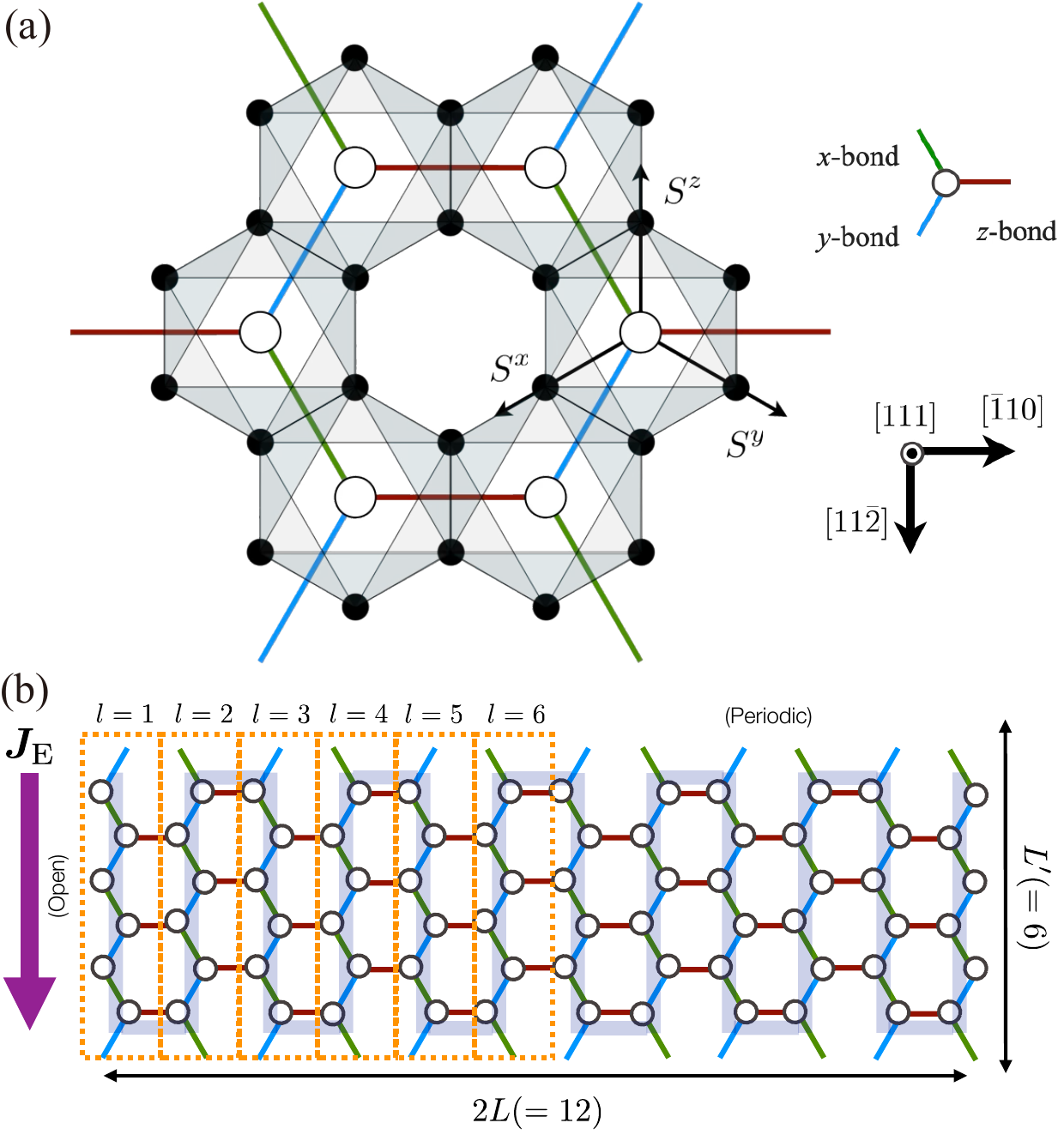}    
  \end{center}
  \caption{
  (a) Schematic view of the extended Kitaev model on the honeycomb lattice in Eq.~\eqref{eq:model-Hamiltonian}. Open circles represent magnetic ions hosting spin-$1/2$ degrees of freedom, while black dots denote nonmagnetic ligands forming an edge-sharing octahedral network.
  The $x$-, $y$-, and $z$-bonds in the Hamiltonian in Eq.~\eqref{eq:model-Hamiltonian} are indicated by green, blue, and red lines, respectively. 
  The spin axes $(S^x, S^y, S^z)$ are also shown, along with the crystallographic directions [$11\bar{2}$], [$\bar{1}10$], and [$111$]. 
  (b) A honeycomb lattice cluster with  $(L, L') = (6, 6) $, consisting of 72 sites, which is typically used in the XTRG calculations. 
  We impose the periodic and open boundary conditions along the vertical and horizontal directions, respectively. 
  We consider the thermal current $\bm{J}_{\mathrm{E}}$ in a downward direction indicated by the purple arrow. 
  Dotted boxes represent the zigzag chains on which the energy current $J_{{\rm E},l}^\parallel$ in Eq.~\eqref{eq:JE} is calculated. 
  In the tensor network simulation, we consider a snake-like matrix product operator indicated by the blue-gray thick line behind the lattice.
  }
  \label{fig:lattice}
\end{figure}

To investigate the thermal Hall conductivity in the Kitaev spin systems, we consider an extended Kitaev model containing the symmetric off-diagonal $\Gamma$ and $\Gamma'$ interactions on the honeycomb lattice. 
The Hamiltonian is given by
\begin{equation}
 \mathcal{H} = \sum_{\gamma = x,y,z} \sum_{\langle i,j \rangle_\gamma}\mathcal{H}_{ij}^\gamma -  \sum_{i,\gamma} h^\gamma S_i^\gamma,
 \label{eq:model-Hamiltonian}
\end{equation}
with
\begin{align}
 \mathcal{H}_{ij}^\gamma &= \Bigl[ K S_i^\gamma S_j^\gamma + \Gamma\left( S_i^\mu S_j^\nu + S_i^\nu S_j^\mu \right) \notag\\
&\qquad + \Gamma'\left( S_i^\mu S_j^\gamma + S_i^\nu S_j^\gamma + S_i^\gamma S_j^\mu + S_i^\gamma S_j^\nu \right) \Bigr]\\
&= \sum_{\alpha,\beta = x,y,z} J_{\alpha\beta}^\gamma S_i^\alpha S_j^\beta, 
\end{align}
where $\langle i,j\rangle_\gamma$ denotes nearest-neighbor pairs on $\gamma$-bonds on the honeycomb lattice [see Fig.~\ref{fig:lattice}(a)], and $S_i^\gamma$ represents the $\gamma$ component of the spin-$1/2$ operator at site $i$; $(\mu, \nu, \gamma)$ is a cyclic permutation of $(x,y,z)$, for example, $(\mu, \nu, \gamma)=(y,z,x)$ for the $x$-bond. 
The second term in Eq.~\eqref{eq:model-Hamiltonian} represents the Zeeman coupling to an external magnetic field $\bm{h} = (h^x, h^y, h^z)$. 
We define the field strength as $h = |\bm{h}| = \sqrt{(h^x)^2+(h^y)^2+(h^z)^2}$. 

A microscopic origin of the Kitaev interaction $K$ in Eq.~\eqref{eq:model-Hamiltonian} was proposed for effective $j_{\rm eff}=1/2$ moments in spin-orbit coupled Mott insulating systems~\cite{Jackeli_PRL2009}. 
It was pointed out that magnetic ions with $t_{2g}^5$ electron configurations comprise the spin-orbit entangled $j_{\rm eff}=1/2$ states under the strong spin-orbit coupling, and exchange processes with indirect hoppings via ligands generate the bond-dependent Kitaev-type interactions between the $j_{\rm eff}=1/2$ moments in edge-sharing octahedral coordinates. 
The symmetric off-diagonal interactions $\Gamma$ and $\Gamma'$ were introduced as additional contributions arising from different exchange processes~\cite{Rau2014}. 
On one hand, the $\Gamma$ interaction is derived by the exchange process including both direct and indirect hoppings, which can be relevant in Kitaev candidate materials~\cite{Yamaji2014,Winter2016,WinterTDBSGV2017,OkuboSYKSTI2017}. 
On the other hand, the $\Gamma'$ interaction is induced by symmetry lowering of the octahedra surrounding magnetic ions from cubic to trigonal symmetry, inevitably existing in quasi-two-dimensional materials~\cite{Rau2014}.
The realistic values of interaction parameters have been extensively discussed for the Kitaev candidate materials~\cite{Yamaji2016,Suzuki2018,laurell2020dynamical,Maksimov2020}.
Most theoretical works have suggested $K<0$ and $\Gamma>0$, with similar magnitudes, for the primary candidate $\alpha$-RuCl$_3$, although the sign of $\Gamma'$ remains under debate.

The $K$-$\Gamma$ model, which is given by Eq.~\eqref{eq:model-Hamiltonian} with $\Gamma^\prime=0$, has been intensively studied as a simplified realistic model, by using various theoretical methods, such as the exact diagonalization (ED)~\cite{catuneanu2018,Yamada2020}, the density matrix renormalization group method~\cite{Gohlke_PRB2018}, the infinite projected entangled pair state method~\cite{Lee_NCom2020,ZhangLLLW2023}, a variational approach~\cite{Zhang2021}, the spin-wave theory~\cite{Smit2020}, and a classical spin approach~\cite{Rayyan2021}.
In the classical limit, this model exhibits macroscopic degeneracy in the ground state. 
Quantum fluctuations lift this degeneracy, but the detailed phase diagram remains controversial.
The introduction of the $\Gamma'$ interaction to the $K$-$\Gamma$ model with $K<0$ and $\Gamma>0$ induces magnetically ordered phases. 
A positive $\Gamma^\prime$ stabilizes a ferromagnetic phase and a chiral spin ordered phase with nonzero spin scalar chirality~\cite{Luo2022PRR,Luo2022}, 
while a negative $\Gamma'$ induces the zigzag order, which has been observed in Kitaev candidate materials at low temperatures~\cite{Rusna2019,gordon2019theory,Chern2020,Lee_NCom2020}. 
Thus, the $K$-$\Gamma$-$\Gamma^\prime$ model in Eq.~\eqref{eq:model-Hamiltonian} might be suitable for investigating the relevant effects of symmetric off-diagonal interactions on the thermal Hall conductivity in real compounds, including $\alpha$-$\mathrm{RuCl_3}$, for which conflicting experimental results have been extensively discussed.

In the following sections, we consider the ferromagnetic Kitaev interaction by setting $K= -1$, which naturally arises from the exchange process in the $t_{2g}^5$ systems{~\cite{Jackeli_PRL2009}}. 
For $\Gamma$ and $\Gamma^\prime$, we consider both positive and negative cases to clarify their effects comprehensively. We take $k_{\rm B} = \hbar = 1$ and set the length of the primitive translation vector of the honeycomb lattice to the unit length.
We typically compute physical quantities for the model on a finite-size cluster with $(L, L') = (6, 6)$ shown in Fig.~\ref{fig:lattice}(b), including $N_{\rm s}=2\times L\times L^{\prime} =72$ spins.

\section{Method}
\label{sec:method}
In this section, we introduce the methods used in this study. 
First, we define the energy current and the thermal Hall conductivity in Sec.~\ref{subsec:Thermal Hall conductivity}. 
Then, we describe two numerical methods for calculating the finite-temperature behaviors: a tensor network method called the XTRG method in Sec.~\ref{subsec:Tensor network method} and the cTPQ state method in Sec.~\ref{subsec:Thermal pure quantum state}. 
We employ the former for main calculations on the $(L, L') = (6, 6)$ cluster introduced above, and use the latter for the benchmark of the XTRG method in smaller size clusters.
Finally, in Sec.~\ref{subsec:Classical Monte Carlo simulation}, we briefly describe the method of classical Monte Carlo simulation to study a classical counterpart to the model.

\subsection{Thermal Hall conductivity}\label{subsec:Thermal Hall conductivity}
\begin{figure}[tbh]
  \begin{center}
    \includegraphics[width=\linewidth]{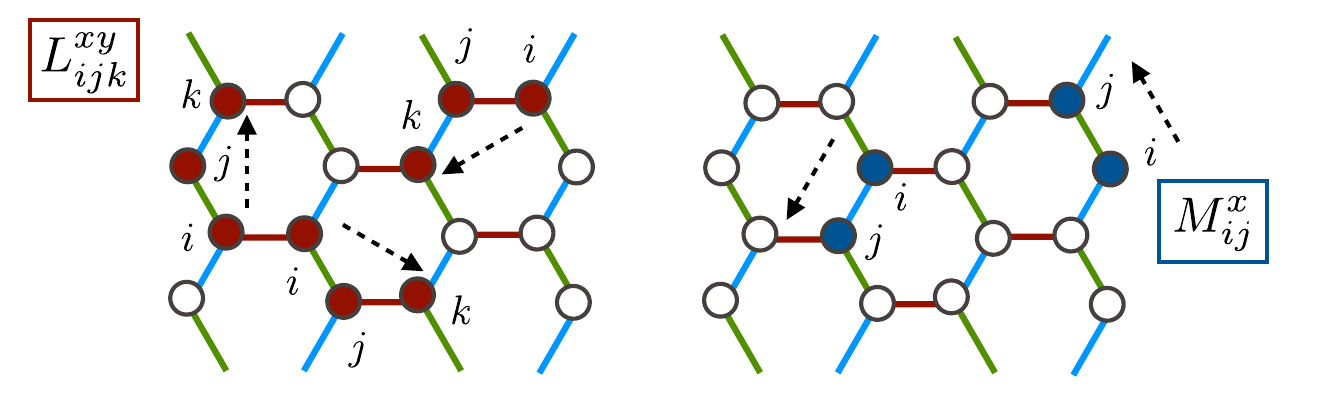}    
  \end{center}
  \caption{
  Graphical representations of the three-spin ($L_{ijk}^{\gamma\gamma^\prime}$) and the two-spin ($M_{ij}^\gamma$) terms in the definition of the thermal current in Eq.~\eqref{eq:def_J}.
  }
  \label{fig:LM_term}
\end{figure}

To investigate the thermal Hall conductivity in Kitaev systems, we define the energy polarization as~\cite{Katsura2010,NasuYM2017}
\begin{equation}
  \bm{P}_{\mathrm{E}} = \sum_{\alpha,\beta,\gamma}\sum_{\langle i,j\rangle_\gamma} \frac{\bm{r}_i + \bm{r}_j}{2} J_{\alpha\beta}^\gamma S_i^\alpha S_j^\beta - \sum_{i,\gamma} \bm{r}_i h^\gamma S_i^\gamma,
\end{equation}
where $\bm{r}_i$ is the position of site $i$. 
From the commutation relation between the Hamiltonian and $\bm{P}_{\mathrm{E}}$, the energy current $\bm{J}_{\mathrm{E}}$ is obtained as 
\begin{align}
  \bm{J}_{\mathrm{E}} &=  i \left[\mathcal{H}, \bm{P}_{\mathrm{E}}\right] \notag \\
&= \sum_{\gamma,\gamma'}\sum_{\langle i,j,k\rangle_{\gamma,\gamma'}} \frac{\bm{r}_k-\bm{r}_i}{2}L_{ijk}^{\gamma\gamma'} + \sum_{\gamma}\sum_{\langle i,j\rangle_{\gamma}} \frac{\bm{r}_j-\bm{r}_i}{2}M_{ij}^{\gamma} 
\label{eq:def_J},
\end{align}
where $\langle i,j,k\rangle_{\gamma,\gamma'}$ represents three neighboring sites consisting of two nearest-neighbor pairs $\langle i,j\rangle_{\gamma}$ and $\langle j,k\rangle_{\gamma'}$ connected at site $j$.
The operator $L_{ijk}^{\gamma\gamma'}$ is the contribution from three-spin correlations,
\begin{equation}
  L_{ijk}^{\gamma\gamma'} = \sum_{\alpha,\beta,\alpha',\beta',\gamma''} J_{\alpha\beta}^\gamma J_{\alpha'\beta'}^{\gamma'} \epsilon_{\alpha\gamma''\alpha'} S_i^\beta S_j^{\gamma''}S_k^{\beta'},
\label{eq:L}
\end{equation}
and $M_{ij}^\gamma$ represents the contribution from two-spin correlations,
\begin{equation}
  M_{ij}^{\gamma} = \sum_{\alpha,\beta,\gamma',\gamma''} J_{\alpha\beta}^\gamma h_{\gamma'} \epsilon_{\gamma'\alpha\gamma''} \left(S_i^{\gamma''} S_j^{\beta} - S_i^{\beta} S_j^{\gamma''} \right),
\label{eq:M}
\end{equation}
where $\epsilon_{\alpha\beta\gamma}$ denotes the completely antisymmetric tensor, which comes from the commutation relation between spins. 
The schematic pictures of these operators are shown in Fig.~\ref{fig:LM_term}. 
It is worth noting that the three-spin product in $L_{ijk}^{\gamma\gamma'}$ coincides with the effective magnetic field derived by third-order perturbations with respect to $\bm{h}$ in the Majorana representation of the Kitaev model~\cite{Kitaev2006,NasuYM2017}.
The effective magnetic field induces the next nearest-neighbor hoppings, which open an excitation gap in the noninteracting Majorana fermion bands and result in nontrivial band topology with nonzero Chern numbers~\cite{Kitaev2006}.
Thus, the $L_{ijk}^{\gamma\gamma'}$ term in Eq.~\eqref{eq:def_J} is expected to contribute to the thermal Hall effect via a chiral edge current, at least, within the pure Kitaev model under a weak magnetic field. 
In contrast, the $M_{ij}^{\gamma}$ term is proportional to the magnetic field, which is not present in the perturbation. 
These two contributions will be examined separately in Sec.~\ref{sec:Results}.

We note that our definition of the energy current is different from that used in Ref.~[\onlinecite{KumarT2023}], where the energy current is defined through the commutation relation of the local Hamiltonians.
The energy current should be defined through the commutation relation of the total Hamiltonian and the total energy polarization as given in Eq.~(\ref{eq:def_J}). 
The energy current used in Ref.~[\onlinecite{KumarT2023}] neglects the position dependence included in the definition of $\vec{P}_{\mathrm{E}}$.
This could explain why $\kappa_{xy}$ in Ref.~[\onlinecite{KumarT2023}] is significantly smaller than the value obtained in this study.

\begin{figure}[t]
  \begin{center}
    \includegraphics[width=\linewidth]{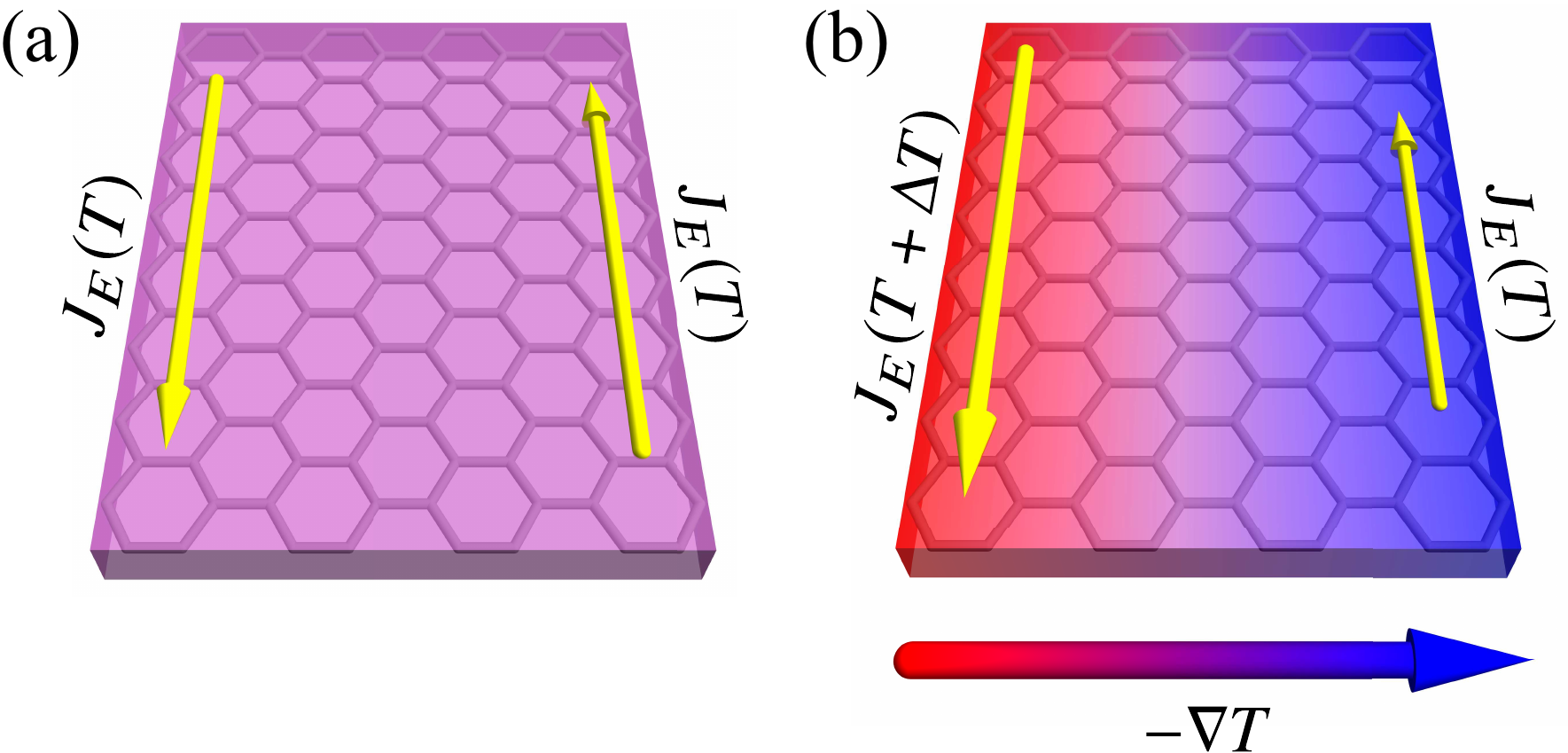}    
  \end{center}
  \caption{
    Schematic view of the energy current (a) without temperature gradient and (b) with temperature gradient. 
    In (a), the energy currents at the edges cancel out, resulting in no net energy flow. 
    In contrast, in (b), the temperature gradient induces a difference in the energy currents, leading to a net energy current transverse to the thermal gradient.
  }
  \label{fig:ThermalHall}
\end{figure}

In the following calculations, we compute the thermal Hall current as the summation of the energy current along the zigzag chains on the honeycomb lattice running in parallel to the open edges. 
The energy current on each zigzag chain, $J_{{\rm E},l}^\parallel$ labeled by $l$, includes the contributions from local currents on the segments inside a box surrounding the chain and across its right edge, which are shown in Fig.~\ref{fig:lattice}(b). 
We compute {$J_{\mathrm{E}}^{\parallel}$} by regarding each term in Eq.~\eqref{eq:def_J} as the current density at $(\bm{r}_i + \bm{r}_j)/2$. Then, we obtain the contribution to the thermal Hall current by summing up $J_{\mathrm{E}}^{\parallel}$ over half of the system from the left edge to the center of the system, which is defined as
\begin{align}
J_{\rm E}^\parallel=\sum_{l=1}^{L}J_{{\rm E},l}^\parallel.
\label{eq:JE}
\end{align} 
Note that the summation over the whole system cancels out at any temperature due to the symmetry in the absence of a temperature gradient [Fig.~\ref{fig:ThermalHall}(a)]. 

To compute the thermal Hall conductivity, we consider a temperature gradient applied perpendicular to the zigzag chains [Fig.~\ref{fig:ThermalHall}(b)]. 
In the limit of a weak temperature gradient, the thermal Hall conductivity on the two-dimensional honeycomb layer is obtained by taking the derivative of $J_{\mathrm{E}}^{\parallel}$ with respect to temperature, 
expressed as 
\begin{equation}
 \kappa_{xy}=\frac{2}{L'} \frac{d \langle J_{\mathrm{E}}^{\parallel}\rangle_{T}}{d T},
\label{eq:def_kxy}
\end{equation}
where $\langle \cdots \rangle_T$ represents the thermal average at temperature $T$.
The coefficient $2/L^\prime$ is required as the thermal Hall conductivity is defined for the energy current per unit area. 
Note that Eq.~\eqref{eq:def_kxy} assumes $J_{\mathrm{E}}^\parallel$ is well localized at the edges. 
This definition of the thermal Hall conductivity is similar to the one used in Ref.~\cite{TangXWT2019}. 
As noted in the literature, the half-quantization of $\kappa_{xy}/T$ is expected to occur within the temperature range of $v/L' \ll T \ll \Delta_{\mathrm{bulk}}$, where $v$ is the velocity of the edge mode and $\Delta_{\mathrm{bulk}}$ is the excitation gap of topological quasiparticles. We will return to this point in Sec.~\ref{subsec:pureKitaev_h111}.

\subsection{Tensor network method}\label{subsec:Tensor network method}
\begin{figure}[t]
  \begin{center}
    \includegraphics[width=\linewidth]{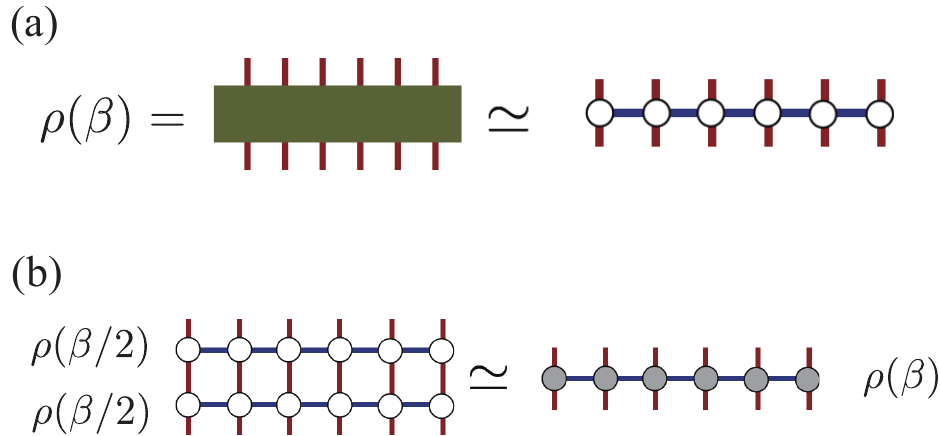}
  \end{center}
  \caption{Tensor network diagram for the density operator approximation. (a) The density matrix is approximated as an MPO with bond dimension $D$. 
  The horizontal line in the right panel corresponds to the gray line in Fig.~\ref{fig:lattice}(b). 
  (b) The density matrix at $\beta$ is calculated as $\rho(\beta)=\rho(\beta/2)\rho(\beta/2)$. 
  The bond dimension of the obtained $\rho(\beta)$ is truncated at $D$ through the standard optimization procedure of MPS.}
  \label{fig:XTRG}
\end{figure}

Let us first introduce the tensor network method employed in our study. 
To calculate finite-temperature properties of the model given by Hamiltonian $\mathcal{H}$, we approximate the density matrix of the system at an inverse temperature $\beta = 1/(k_{\mathrm{B}} T)$, $\rho(\beta) = e^{-\beta\mathcal{H}}$, by a matrix product operator (MPO) with bond dimension $D$ [see Fig.~\ref{fig:XTRG}(a)]. In the actual calculations, the string of the MPO is arranged in a snake form on the honeycomb cluster, as indicated by the blue-gray line in Fig.~\ref{fig:lattice}(b). 

To optimize the tensors in an MPO as the density matrix at $\beta$, 
we employ the XTRG approach~\cite{Chen2018}, which has been successfully used to calculate finite-temperature properties of extended Kitaev models~\cite{Li2020, LiZWWGQLGL2021}. 
In the XTRG method, we calculate the density matrix at $\beta$ through the relationship
\begin{equation}
  \rho(\beta)=\rho(\beta/2)\rho(\beta/2).
\end{equation}
When $\rho(\beta/2)$ is represented by an MPO with the bond dimension $D$, $\rho(\beta)$ becomes an MPO with the bond dimension $D^2$. 
We approximate $\rho(\beta)$ by an MPO with bond dimension $D$ through the standard optimization procedure for the matrix product states (MPS) \cite{Chen2018} [see Fig.~\ref{fig:XTRG}(b)]. 
In the optimization, we employ the two-site update, resulting in the total computational cost that scales as $O(D^4)$. 

As the initial condition of the XTRG method, we prepare $\rho(\beta_0)$ with $\beta_0 = 10^{-7}$ through the approximated form $\rho(\beta_0) \simeq 1 - \beta_0\mathcal{H}$, where the Hamiltonian is represented as an MPO. 
We calculate the expectation value of an operator ${O}$ through thus obtained $\rho(\beta)$ as $\langle {O} \rangle = \mathrm{Tr}[{O}\rho] /\mathrm{Tr}\rho$ (note that the density matrix is not normalized). 
The temperature derivative of $\langle {O} \rangle$ is computed by numerical differentiation.
In the following, we mainly show the data with $D=500$ for the $(L, L') = (6, 6)$ cluster. 
Benchmark calculations for the values of $D$ and the cluster size and geometry are shown in Appendix~\ref{app:XTRG_Bench}. 

\subsection{Thermal pure quantum state}\label{subsec:Thermal pure quantum state}
Next, we describe the cTPQ state method~\cite{Sugiura_PRL2013}, which enables us to calculate finite-temperature properties of quantum many-body systems using the power method. We note that several similar methods were independently proposed~\cite{Imada_JPSJ1986,Jaklic_PRB1994,Hams_PRE2000,Lloyd} before the proposal of the cTPQ state method~[\onlinecite{Sugiura_PRL2013}].
The cTPQ state is constructed as
\begin{align}
  |\Phi_{\rm cTPQ}^{p}(\beta)\rangle = \rho(\beta/2) |\Phi_{\rm rand}^{p}\rangle,
\end{align}
where $|\Phi_{\rm rand}^{p}\rangle$ is the $p$th initial random vector, which is uniformly distributed on the $N_{\rm H}$-dimensional hypersphere ($N_{\rm H}$ is the dimension of the Hilbert space of the given system).
Any local physical quantity at inverse temperature $\beta$ can be calculated as the expectation values of $|\Phi_{\rm cTPQ}^{p}(\beta)\rangle$, i.e.,
\begin{align}
\langle A(\beta)\rangle
=\frac{\langle \Phi_{\rm cTPQ}^{p}(\beta)|A|\Phi_{\rm cTPQ}^{p}(\beta)\rangle}
{\langle \Phi_{\rm cTPQ}^{p}(\beta)|\Phi_{\rm cTPQ}^{p}(\beta)\rangle}.
\end{align}
We numerically obtain the cTPQ state by 
\begin{align}
\rho(\beta/2) |\Phi_{\rm rand}^{p}\rangle=U(\Delta\tau)^{k}|\Phi_{\rm rand}^{p}\rangle,
\end{align}
where $\beta=k\Delta\tau$ and 
\begin{align}
U(\Delta\tau)=\exp
\left(-\frac{\Delta\tau}{2}\mathcal{H}\right)
\simeq\sum_{n=0}^{n_{\rm max}}\frac{1}{n!}\left(-\frac{\Delta\tau}{2}\mathcal{H}\right)^{n}. 
\label{eq:U_Deltatau}
\end{align}

The cTPQ state method provides the numerically exact
results within the statistical errors, which are defined by the statistical distribution of the initial random vectors.
In actual calculations, we use $\mathcal{H}\Phi$~\cite{Kawamura_CPC2017,HPhi_v2,HPhi_release}, taking $n_{\rm max}=6$ and $\Delta\tau =0.02$ in Eq.~\eqref{eq:U_Deltatau}.
To estimate the statistical errors of the cTPQ method, we employ the bootstrap method~\cite{HPhi_v2}.
Benchmark results of the cTPQ state method, in comparison with the ED results, are shown in Appendix~\ref{app:cTPQ}. 
In this study, we use the cTPQ state method to validate the reliability of the XTRG method. 
The benchmark results are presented in Appendix~\ref{app:XTRGcTPQ}.

\subsection{Classical Monte Carlo simulation}\label{subsec:Classical Monte Carlo simulation}
Finally, we describe the method of classical Monte Carlo simulation.
In this method, an $S=1/2$ spin at each site is regarded as a classical vector with length $1/2$.
The classical spin at site $i$ is parameterized by $\theta_i$ and $\phi_i$ as $\bm{S}_i = \frac{1}{2}(\sin\theta_i\cos\phi_i, \sin\theta_i\sin\phi_i, \cos\theta_i)$.
In the calculations of the thermal average, the integral $\int \prod_i d\phi_i d\theta_i \sin\theta_i$ is evaluated using the Markov-chain Monte Carlo method.
To accelerate computations and avoid trapping spin configurations in local minima, we use the replica exchange method \cite{Hukushima1996}.
In the simulations, we prepare 48 replicas with different temperatures.
We perform 10~000~000~MC steps for measurements after 10~000~MC steps for thermalization in the 800-site cluster with $(L, L') = (10, 40)$. 
The temperature derivative of $\langle O \rangle$ is evaluated by the correlation with ${\cal H}$ as  
\begin{align}
\frac{d\langle O \rangle}
{dT}=\frac{
\langle O 
{\cal H}\rangle -\langle O 
\rangle \langle {\cal H}\rangle} 
{T^2},  
\end{align}
which gives accurate estimates compared to numerical differentiation. 

\section{Result}\label{sec:Results}
In this section, we present the results of our numerical simulations. First, we discuss the pure Kitaev model in Sec.~\ref{sec:pureKitaev}, focusing on the case of the magnetic field applied along the [$111$] direction, as well as the dependence on the field direction. 
Then, we examine the effects of the symmetric off-diagonal interactions $\Gamma$ and $\Gamma'$ in Sec.~\ref{sec:Gamma}. In Sec.~\ref{sec:classicalMC}, we compare the results of the extended Kitaev model with those of its classical counterpart, obtained by classical Monte Carlo simulations. 
Finally, in Sec.~\ref{sec:summary_of_temp_field}, we summarize the temperature and field dependence of the thermal Hall conductivity to highlight the effects of $\Gamma$ and $\Gamma'$.

\subsection{Pure Kitaev model}
\label{sec:pureKitaev}
\subsubsection{Magnetic field along [111] direction}\label{subsec:pureKitaev_h111}
\begin{figure}[tbh]
  \begin{center}
    \includegraphics[width=\linewidth]{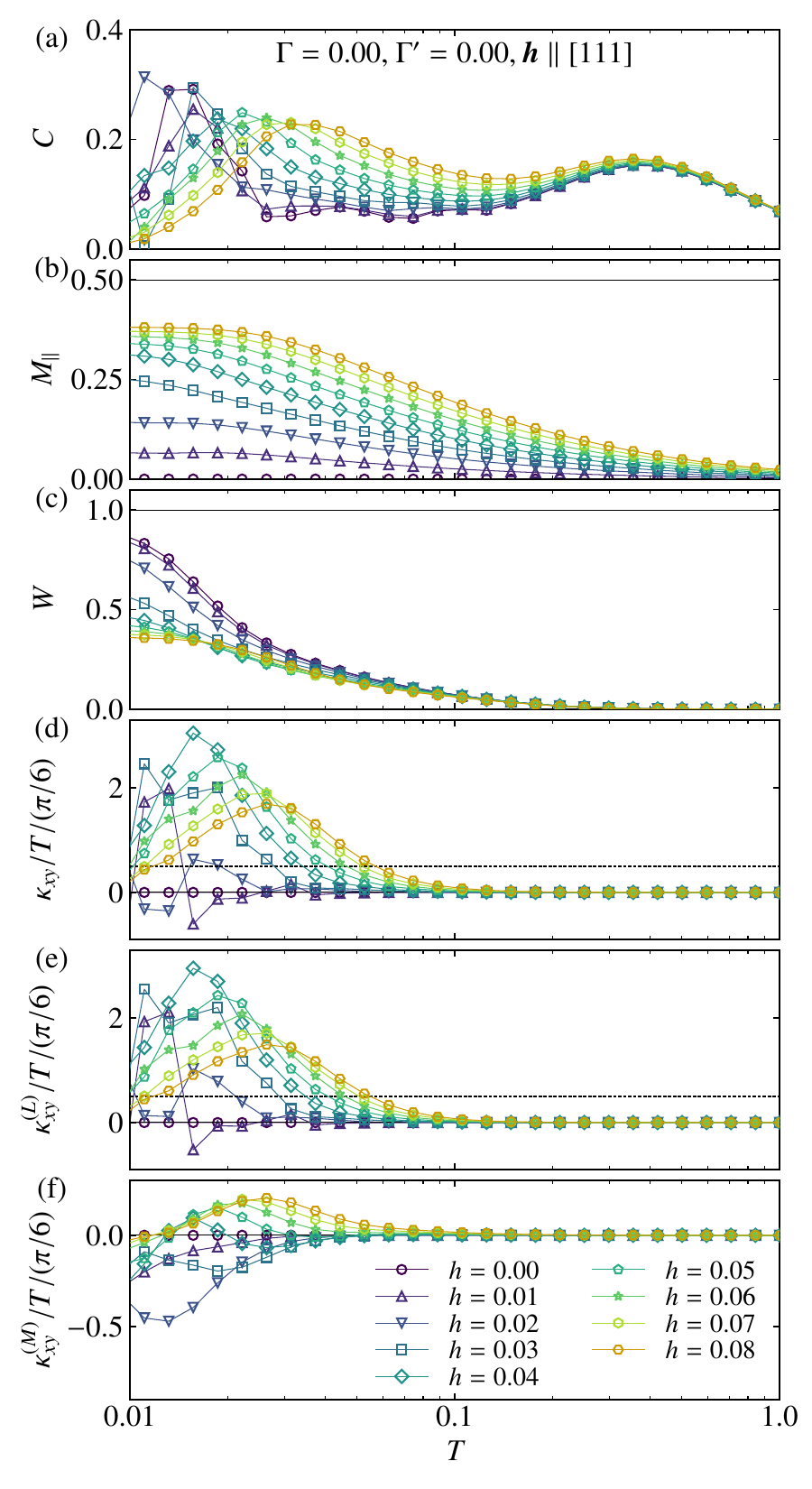}
  \end{center}
  \caption{Temperature dependence of (a) the specific heat [Eq.~\eqref{eq:C}], (b) the magnetization [Eq.~\eqref{eq:Mag}], (c) the flux density [Eq.~\eqref{eq:W}], and (d) the thermal Hall conductivity $\kappa_{xy}$ [Eq.~\eqref{eq:def_kxy}] divided by $T$ of the pure Kitaev model for several values of the external magnetic field parallel to the $[111]$ direction ($h=|\bm{h}|$). (e), (f) Contributions from the three-spin term in Eq.~\eqref{eq:L} and the two-spin term in Eq.~\eqref{eq:M} to $\kappa_{xy}/T$, respectively. The data in (d), (e), and (f) are plotted in units of $\pi/6$, and the dashed horizontal line in each figure indicates the half-integer quantized value $\pi/12$.
  }
  \label{fig:CMF_pure}
\end{figure}

We begin with the model with only the ferromagnetic Kitaev interaction ($K=-1$ and $\Gamma = \Gamma^\prime = 0$) in an applied magnetic field along the $[111]$ direction ($h^x=h^y=h^z=h/\sqrt{3}$). 
In this setup, we expect $\kappa_{xy}/T$ to be quantized at a positive value of $\pi/12$ for small magnetic fields in the zero temperature limit~\cite{Kitaev2006}. 
Figures~\ref{fig:CMF_pure}(a)-\ref{fig:CMF_pure}(c) show the temperature dependence of the specific heat $C$, the magnetization along the magnetic field $M_{\parallel}$, and the flux density $W$, respectively given by
\begin{align}
C&=\frac{\ev*{{\cal H}^{2}}-\ev*{{\cal H}}^2}{T^2}, \label{eq:C}\\
M_{\parallel}&=\ev*{\vec{M}}
\cdot\frac{\vec{h}}{|\vec{h}|},~~\vec{M}=\frac{1}{N_{\rm s}}\sum_{i}\vec{S}_{i}, \label{eq:Mag}\\
W&=\frac{1}{N_{\rm h}}\sum_{p}\ev*{W_{p}}, 
~~W_{p}=2^6\prod_{i\in p}S^{\gamma_{i}}_{i} \label{eq:W},
\end{align}
where $p$ runs over all hexagons in the honeycomb lattice and $N_{\rm h}$ is the number of hexagons; $\gamma_{i}$ denotes the bond component that does not belong to the edges of $p$ at site $i$.

We first discuss the temperature dependence of the specific heat $C$.
As shown in Fig.~\ref{fig:CMF_pure}(a), clear double-peak structures are observed at each magnetic field.
These features evolve smoothly from the characteristic behavior of the Kitaev model at zero field, which arises from fractionalization of spins into itinerant Majorana fermions and localized $Z_2$ fluxes~\cite{NasuUM2014,NasuUM2015}.
As the magnetic field increases, the low-temperature peak shifts to higher temperatures, while the high-temperature peak remains largely unaffected. 
This shift of the low-temperature peak is not observed in the calculations where the magnetic field is introduced perturbatively~\cite{NasuYM2017}, highlighting the importance of non-perturbative effects.

In the low-field region for $h\lesssim 0.03$, the specific heat exhibits non-smooth temperature dependence at low temperatures. 
This behavior may originate from the finite bond dimension $D$ used in the XTRG calculations. 
In contrast, in the higher-field region for $h \gtrsim 0.04$, the specific heat exhibits smooth temperature dependence down to $T=0.01$. 
This suggests that $D=500$ is sufficient to accurately capture the thermal properties of the Kitaev model in this field region. 
Further discussions on the $D$ dependence are shown in Appendix~\ref{app:XTRG_Bench}. 

Despite the non-smooth behavior at low fields and low temperatures, the magnetic-field dependence of the low-temperature peak appears to be consistent with the previous studies~\cite{YoshitakeNKM2020,Li2020,LiZWWGQLGL2021,LiLXGQLS2024}. 
Specifically, the peak shifts to lower temperatures under weak magnetic fields, and moves to higher temperatures as the magnetic field increases beyond $h \gtrsim 0.02$.
This behavior is likely associated with the quantum phase transition in the Kitaev model, which separates a low-field topological chiral spin liquid from a high-field polarized state at $h_c \simeq 0.024$~\cite{Gohlke2018}. 

\begin{figure}[tbh]
  \begin{center}
    \includegraphics[width=\linewidth]{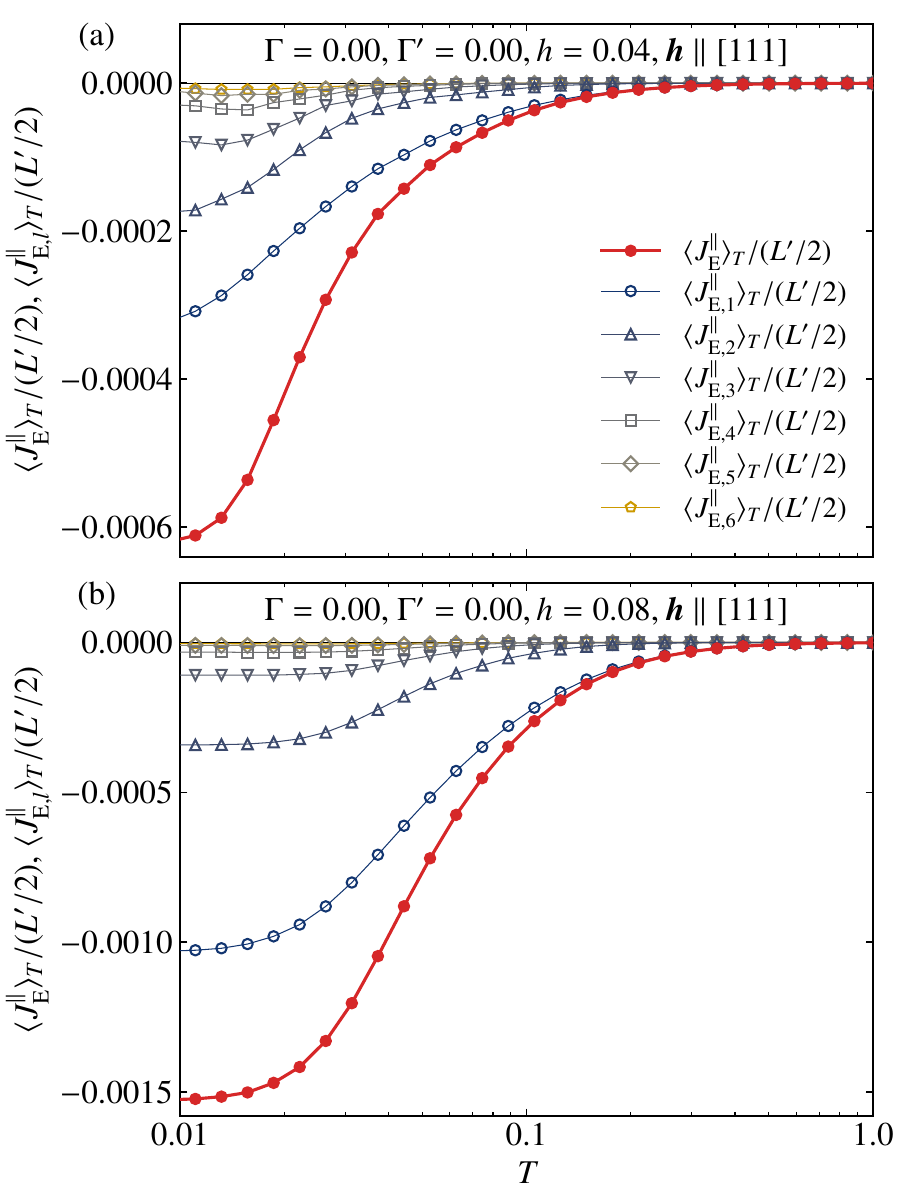}
  \end{center}
  \caption{Temperature dependence of the energy current of the pure Kitaev model at (a) $h=0.04$ and (b) $h=0.08$. In addition to the total energy current (red filled symbols), the contribution from each line is shown (colored open symbols); see Eq.~\eqref{eq:JE}.
}
  \label{fig:J_line}
\end{figure}

\begin{figure}[tbh]
  \begin{center}
    \includegraphics[width=\linewidth]{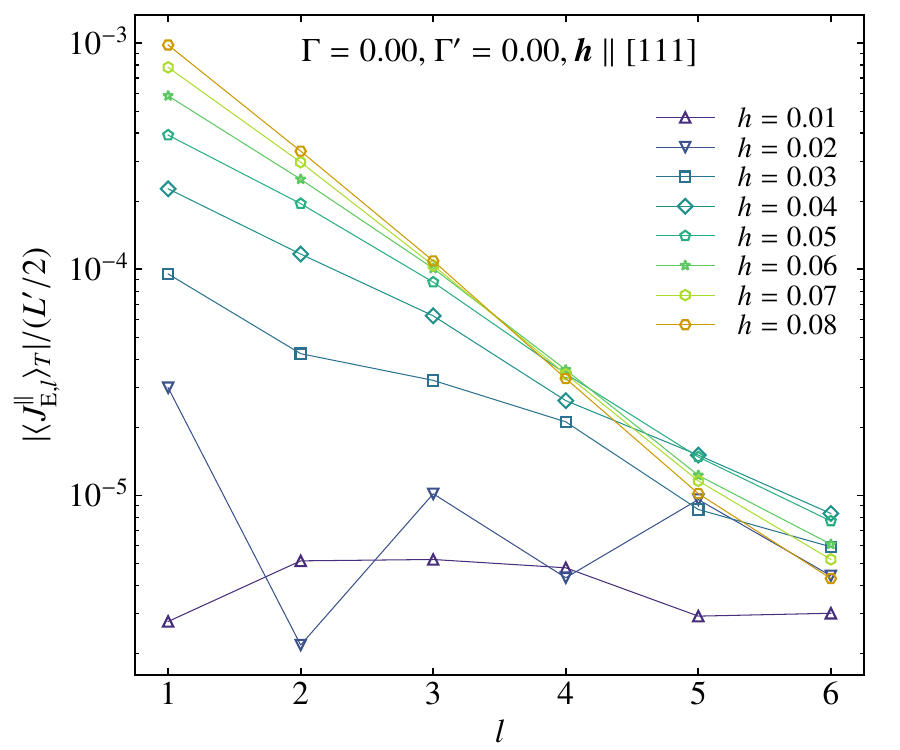}
  \end{center}
  \caption{Semi-log plot of the position ($l$) dependence of the energy current amplitude for the pure Kitaev model at several values of the external magnetic field parallel to the $[111]$ direction at $T \simeq 0.0186$.} 
  \label{fig:J_line_dep}
\end{figure}

Next, we discuss the temperature dependence of the magnetization $M_\parallel$ and the flux density $W$ shown in Figs.~\ref{fig:CMF_pure}(b) and \ref{fig:CMF_pure}(c), respectively. 
As expected, $M_\parallel$ monotonically increases with the magnetic field, reaching over 60\% of the saturation value of $0.5$ for $h\gtrsim 0.05$ at low temperatures. 
Since the applied magnetic field inducing nonzero magnetization renders the flux operator on each hexagonal plaquette a nonconserved quantity, $W$ decreases monotonically with increasing magnetic field, down to $\sim 0.4$ for $h\gtrsim 0.05$ at low temperatures, which is significantly lower than $W=1$ expected for the ideal Kitaev QSL.
We note that both $M_{\parallel}$ and $W$ decrease rapidly around at $h_c \simeq 0.024$, suggesting a signature of the quantum phase transition between the topological and polarized phases. 

Figure~\ref{fig:CMF_pure}(d) displays the temperature dependence of the thermal Hall conductivity divided by temperature, $\kappa_{xy}/T$. 
As this is obtained by the temperature derivative of the energy current through Eq.~\eqref{eq:def_kxy} with Eq.~\eqref{eq:JE}, let us discuss the behavior of the energy current first. 
Figure~\ref{fig:J_line} shows the temperature dependence of the total energy current $\langle J_{\mathrm{E}}^\parallel \rangle_T$ defined in Eq.~\eqref{eq:JE} at two representative magnetic fields, $h=0.04$ and $h=0.08$ (red filled symbols). 
We find that $\langle J_{\mathrm{E}}^\parallel \rangle_T$ monotonically decreases while decreasing temperature, starting from $\langle J_{\mathrm{E}}^{\parallel} \rangle_T =0$ in the high-temperature limit. This behavior indicates that $\kappa_{xy}$ is positive in the entire temperature range, which is consistent with the expectation in the zero temperature limit. 
In addition, $\langle J_{\mathrm{E}}^\parallel \rangle_T$ exhibits an inflection point at an intermediate temperature, indicating that $\kappa_{xy}$ has a peak at the corresponding temperatures, as discussed below. 
Notably, the inflection point appears to correlate with the low-temperature peak of the specific heat in Fig.~\ref{fig:CMF_pure}(a). 

Figure~\ref{fig:J_line} also plots the contributions from each zigzag chain, $\langle J_{\mathrm{E},l}^\parallel \rangle_T$. 
We observe that these contributions exhibit similar temperature dependence, while they decrease rapidly with increasing distance from the edge, $l$. 
To highlight this behavior more clearly, we display the $l$ dependence of the energy current amplitude at several magnetic fields for $T\simeq 0.0186$ in Fig.~\ref{fig:J_line_dep}. 
We find that the amplitude of the energy current is largest at the edge ($l=1$), and decreases almost exponentially toward the center of the system ($l=6$) [see Fig.~\ref{fig:lattice}(b)]. 
This result demonstrates that the energy current is dominated by the edge current, similar to the chiral edge current originating from the topologically-nontrivial Majorana band structure predicted by the perturbation theory.
The contributions become negligibly small near the center, suggesting that the employed system size, $L=6$, is large enough to capture the essential features of this edge current.
Surprisingly, similar exponential decays are also observed at higher temperatures as shown in Appendix~\ref{app:J_line}. This supports the assumption made in Eq.~\eqref{eq:def_kxy}, 
across a broad range of temperatures and magnetic fields in our calculations.

It is worth noting that the edge current is nonzero even at high fields well beyond $h_c \simeq 0.024$, i.e., within the polarized regime beyond the quantum critical point. 
Moreover, the current amplitude increases with the field and decays more rapidly toward the center. 
Although this rapid decay may be attributed to the increase in the excitation gap induced by the Zeeman coupling, the nonzero and growing current amplitude suggests the presence of nontrivial quasiparticles even within the polarized regime.

Based on these results, let us now discuss the temperature dependence of $\kappa_{xy}/T$ shown in Fig.~\ref{fig:CMF_pure}(d). First of all, $\kappa_{xy}/T$ is overall positive, except for small $h$ and low $T$, and exhibits a nonmonotonic temperature dependence with a clear peak structure, which shifts to higher temperatures with increasing $h$. 
The peak temperature exhibits a correlation with the low-temperature peak of the specific heat in Fig.~\ref{fig:CMF_pure}(a).
These behaviors are expected from $\langle J_{\mathrm{E}}^\parallel \rangle_T$ in Fig.~\ref{fig:J_line}. 
Surprisingly, the peak value of $\kappa_{xy}/T$ largely exceeds the half-integer quantized value indicated by the broken line in Fig.~\ref{fig:CMF_pure}(d). 
This overshooting behavior becomes more pronounced with increasing $h$, reaching a maximum at $h\simeq 0.04$, and then gradually decreases at higher $h$. 
Such behavior was not observed in the previous numerical calculations, which treated the magnetic fields as effective three-spin interactions~\cite{NasuYM2017} or employed a simplified definition of the energy current~\cite{KumarT2023}. 

Thus, our results indicate that full quantum calculations, using a proper definition of the energy current, successfully reproduce the overshooting behavior observed in a wide range of magnetic fields in experiments. 
This finding suggests that the interesting feature can be attributed solely to magnetic origins, without considering contributions from phonons~\cite{Ye2018Quantization,Vinkler2018}. 
Similar overshooting behavior was also discussed in the context of topological vison bands, which arise only in the system with antiferromagnetic Kitaev coupling~\cite{Joy2022}. 
However, our result is for the ferromagnetic Kitaev coupling, which is realistic for Kitaev magnets, such as $\alpha$-RuCl$_3$, indicating a different underlying mechanism from that of vison bands. 

To gain deeper insight into the origin of this peculiar behavior, we decompose it into two components, $\kappa_{xy}^{(L)}$ and  $\kappa_{xy}^{(M)}$, which correspond to the contributions from three-spin and two-spin correlations defined in Eqs.~(\ref{eq:L}) and (\ref{eq:M}), respectively.
Figures~\ref{fig:CMF_pure}(e) and \ref{fig:CMF_pure}(f) show the temperature dependence of $\kappa_{xy}^{(L)}$ and  $\kappa_{xy}^{(M)}$, respectively.
We find that the three-spin contribution to $\kappa_{xy}^{(L)}/T$ dominates the total $\kappa_{xy}/T$ across all magnetic fields.
In contrast, the two-spin contribution $\kappa_{xy}^{(M)}/T$ is an order of magnitude smaller than that from the three-spin part and plays a minor role.
As mentioned in Sec.~\ref{subsec:Thermal Hall conductivity}, the three-spin terms in Eq.~\eqref{eq:L} can be interpreted as an effective magnetic field arising from the third-order perturbations in $\bm{h}$ within the Majorana fermion representation, while the two-spin terms do not appear in perturbation theory. 
The fact that the induced energy current is well localized at the edge further supports the scenario for the three-spin terms. 
These observations collectively suggest that $\kappa_{xy}^{(L)}$ predominantly reflects contributions from the topological gap opening in the emergent Majorana fermion system, whereas
$\kappa_{xy}^{(M)}$ may include effects beyond the Majorana fermion picture.
Given that $\kappa_{xy}$ is dominated by $\kappa_{xy}^{(L)}$, 
its behavior in the calculated field range can be reasonably interpreted within the topological Majorana fermion picture.

One may notice that $L_{ijk}^{\gamma\gamma'}$ defined in Eq.~\eqref{eq:L} is closely related to 
the scalar spin chirality, $\bm{S}_i\cdot (\bm{S}_j \times \bm{S}_k)$, for the pure Kitaev model, where $J^\gamma_{\alpha\beta} \propto \delta_{\alpha\beta}$. 
Since the scalar spin chirality can become nonzero in a chiral spin liquid under a magnetic field, it may contribute to the bulk thermal Hall conductivity via $L_{ijk}^{\gamma\gamma'}$.
However, our results indicate that this bulk contribution is negligibly small compared to the dominant edge contributions discussed above. 

It is worth emphasizing that our results of $\kappa_{xy}/T$ exhibit a pronounced peak at intermediate temperatures across a wide range of magnetic fields, including the polarized regime well beyond the critical field $h_c \simeq 0.024$. 
In this high-field polarized regime, the conventional magnon picture is typically considered more appropriate than the topological Majorana fermion picture. 
Indeed, thermal Hall transport in this regime was discussed using linear-spin wave theory based on the conventional magnon picture~\cite{McClartyDGRPMP2018}. 
This approach concluded that $\kappa_{xy}/T$ becomes nonzero due to the topological properties of magnons, showing a sign change from positive to negative with increasing temperature. 
A subsequent study, which includes magnon-magnon interactions beyond the linear spin-wave theory, demonstrated that $\kappa_{xy}/T$ is suppressed and becomes consistently negative in the field range considered in our calculations~\cite{Koyama2024}. 
These results are in stark contrast to our results, which not only differ in sign but also align with experimental observations. 

In the Kitaev model under an applied magnetic field, it is expected that the system undergoes a crossover from the Majorana fermion picture, which gives an exact solution at zero field, to the magnon picture, which becomes valid in the strong field limit. 
However, our results indicate that the topological Majorana fermion picture remains relevant across a broad range of magnetic fields, extending well beyond the quantum phase transition. 
Notably, this conclusion is consistent with previous quantum Monte Carlo results~\cite{YoshitakeNKM2020}, further supporting the robustness of the topological Majorana fermion picture over a wide field range. 

Finally, let us discuss the asymptotic half-integer quantization at low temperatures. 
We note that while $\kappa_{xy}/T$ takes a value around the half-integer quantization $\pi/12$ at the lowest temperature calculated, it does not exhibit clear convergence to this value, even when the magnetic field is sufficiently small.
This is presumably due to multiple reasons, including the finite bond dimension $D$, finite-size effects, and insufficiently low temperatures.
First, as discussed for the specific heat above, the bond dimension $D=500$ is not sufficient to yield reliable results at low temperatures for $h\lesssim0.04$.
Second, finite-size effects become more pronounced at smaller $h$, where the excitation gap is expected to be smaller. 
This leads to a broader spatial extension of the edge current into the bulk, causing overlap between contributions from both edges in the finite-size cluster and thereby hampering accurate estimation of the edge current. 
Moreover, the finite length of the cluster along the edges, $L'$, further contributes to deviations from quantization by discretizing the chiral edge mode; clear quantization is expected only when $T \gg v/L'$~\cite{TangXWT2019}.
Lastly, asymptotic quantization is expected only at temperatures much lower than the topological bulk gap, i.e., when $T \ll \Delta_{\mathrm{bulk}}$~\cite{TangXWT2019}. 
However, the precise value of this gap is not known beyond perturbation theory. 
Addressing these issues requires more sophisticated numerical methods capable of accessing larger-size systems at lower temperatures. 

\subsubsection{Field-direction dependence}
\label{subsec:field_direction_dep}

\begin{figure}[tbh]
  \begin{center}
    \includegraphics[width=\linewidth]{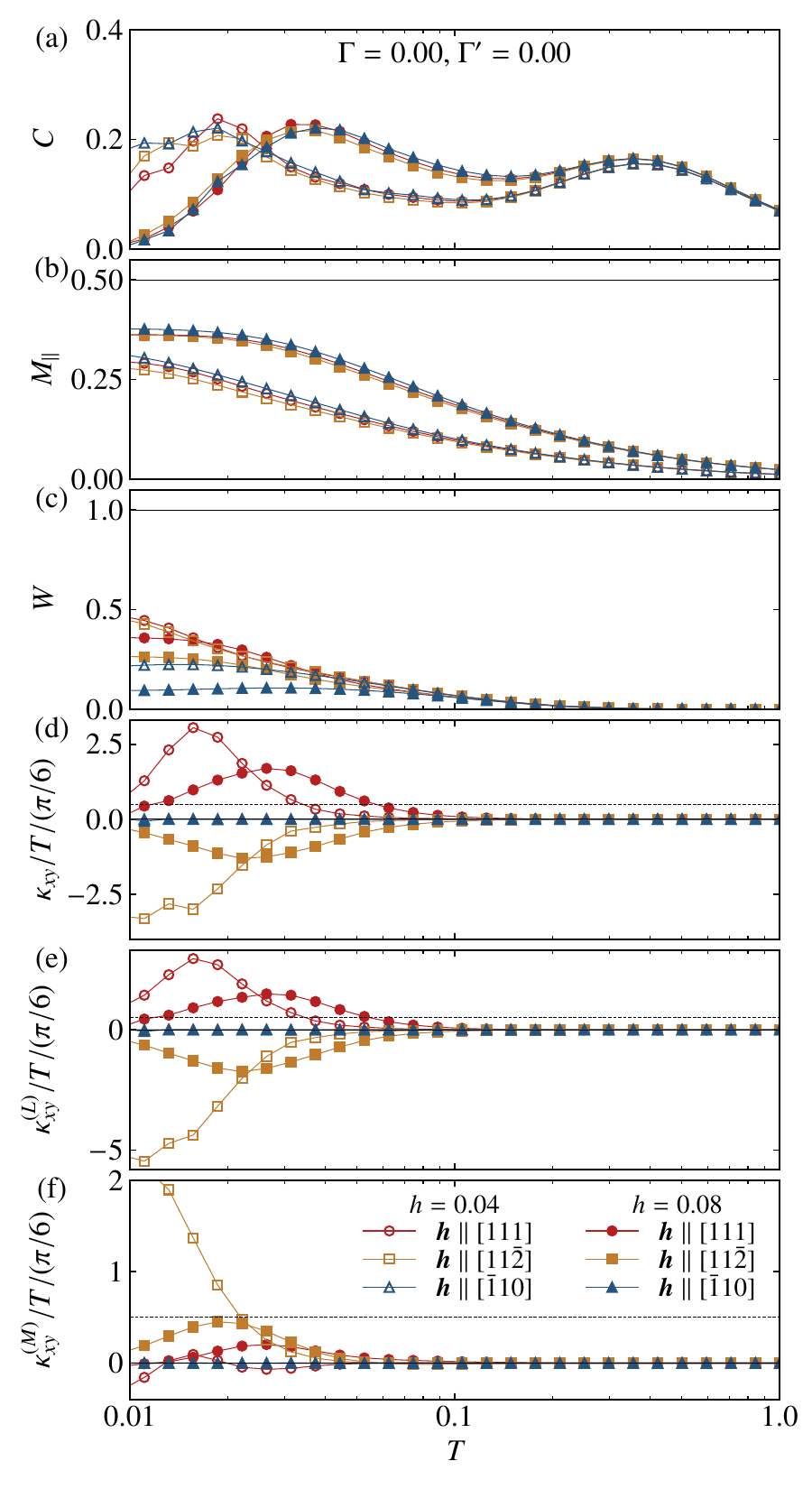}
  \end{center}  
  \caption{Temperature dependence of (a) the specific heat, (b) the magnetization, (c) the flux density, (d) $\kappa_{xy}/T$, and (e) three-spin and (f) two-spin contributions to $\kappa_{xy}/T$ of the pure Kitaev model under magnetic fields parallel to $[111]$, $[11\bar{2}]$, and $[1\bar{1}0]$ with $h=0.04$ and $h = 0.08$.
  }
  \label{fig:CMF_ab}
\end{figure}
In this section, we examine the dependence of physical quantities on the direction of the magnetic field in the pure Kitaev model.
Specifically, we consider three field directions:
[$111$], [11$\bar{2}$], and [$\bar{1}$10]. 
Note that the data for the [$111$] direction are the same as those presented in the previous section. 
When the spin axes $(S^x, S^y, S^z)$ are aligned wth the cartesian coordinates defined by the directions connecting the magnetic ion and the surrounding ligands in the edge-sharing honeycomb network of octahedra~\cite{Jackeli_PRL2009},  the [$111$] direction corresponds to out-of-plane, while the [11$\bar{2}$] and [$\bar{1}$10] directions lie in the plane [see Fig.~\ref{fig:lattice}(a)]. 
The perturbation theory predicts that the topology in the Majorana fermion bands changes with the field direction: a gap in the Majorana excitations opens in proportion to the product of three field components, $h^x h^y h^z$, and the Chern number $\nu$ changes its sign depending on the sign of $h^x h^y h^z$~\cite{Kitaev2006}. 
Consequently, $\kappa_{xy}/T$ in Eq.~\eqref{eq:k_xy} is expected to change its sign with the sign of $h^x h^y h^z$ in the weak field limit. 
Accoding to this prediction, $\kappa_{xy}/T$ should be positive for the [$111$] field, negative for the [$11\bar{2}]$ field, and vanishes for the [$\bar{1}10$] field.
We test these predictions beyond the perturbative regime by performing numerical calculations.

Figure~\ref{fig:CMF_ab} shows the results for  $h=0.04$ and $h=0.08$. 
We find that the specific heat and the magnetization exhibit no significant changes across different field directions, as shown in Figs.~\ref{fig:CMF_ab}(a) and \ref{fig:CMF_ab}(b), respectively. 
In contrast, the flux density shows a significant decrease for the [$\bar{1}$10] direction, as shown in Fig.~\ref{fig:CMF_ab}(c). 
This behavior may be associated with the presence of low-energy excitations, consistent with the prediction from the perturbation theory that the Majorana excitation gap does not open for a magnetic field applied in this direction~\cite{Kitaev2006}.

The thermal Hall conductivity exhibits strong dependence on the direction of the applied magnetic field, consistent with predictions from the perturbation theory.
As shown in Fig.~\ref{fig:CMF_ab}(d), we find that $\kappa_{xy}/T$ reverses its sign from positive to negative when the field direction is changed from [$111$] to  [$11\bar{2}$]. 
We also find that the three-spin contribution $\kappa_{xy}^{(L)}$ dominates in both cases and is responsible for the observed sign change of the total $\kappa_{xy}$.
In addition, $\kappa_{xy}/T$ is vanishingly small for the [$\bar{1}11$] field. 
These results suggest that the topological Majorana fermion picture provides a plausible explanation for the field-direction dependence of the thermal Hall effect, even in the wide range of magnetic field strengths beyond the perturbative region. 

We again emphasize that these behaviors persist even at field strengths beyond the critical field value, $h_c \simeq 0.024$. 
Although results based on the conventional magnon picture are not readily available for the [$11\bar{2}$] and [$\bar{1}11$] field directions, our findings consistently support the relevance of topological Majorana fermions in the enhanced thermal Hall response, even within the polarized regime beyond the quantum critical point. 

We note that the vanishing of $\kappa_{xy}/T$ for the [$\bar{1}10$] field at all temperatures can be understood from the symmetry of the Hamiltonian.
The Hamiltonian in Eq.~\eqref{eq:model-Hamiltonian} is invariant under the following combined operations: 
a $C_2$ rotation in spin space about the $[\bar{1}10]$ axis, which transforms $(S^x,S^y,S^z)$ to $(-S^y,-S^x,-S^z)$, and a rotation of the honeycomb plane in real space about the $z$~bond, which exchanges the $x$ and $y$~bonds.
This real-space rotation of the honeycomb plane inverts the component of the position vector $\bm{r}$ along the zigzag edge, and thereby, the thermal current along the zigzag edge should be zero.
Note that this symmetry argument remains valid even in the presence of the $\Gamma$ and $\Gamma'$ interactions, as well as in the classical case.

\subsection{Effect of non-Kitaev interactions}
\label{sec:Gamma}
In this subsection, we investigate the effects of the symmetric off-diagonal interactions, the $\Gamma$ and $\Gamma'$ terms, on the thermal Hall conductivity.
As introduced in Sec.~\ref{sec:introduction}, the perturbation theory in the Majorana fermion representation predicts that a negative $\Gamma^{\prime}$ enhances the Majorana gap~\cite{TakikawaF2020}, and hence, the thermal Hall conductivity. 
In contrast, $\Gamma$ appears at higher-order perturbations, and its effect on the thermal Hall conductivity remains unclear.
In the following, we examine these predictions beyond the perturbative regime by numerics, focusing on the system under an applied magnetic field along the $[111]$ direction.

\subsubsection{Symmetric off-diagonal interaction $\Gamma^{\prime}$}\label{subsec:Gamma_prime}
\begin{figure}[tbh]
  \begin{center}
    \includegraphics[width=\linewidth]{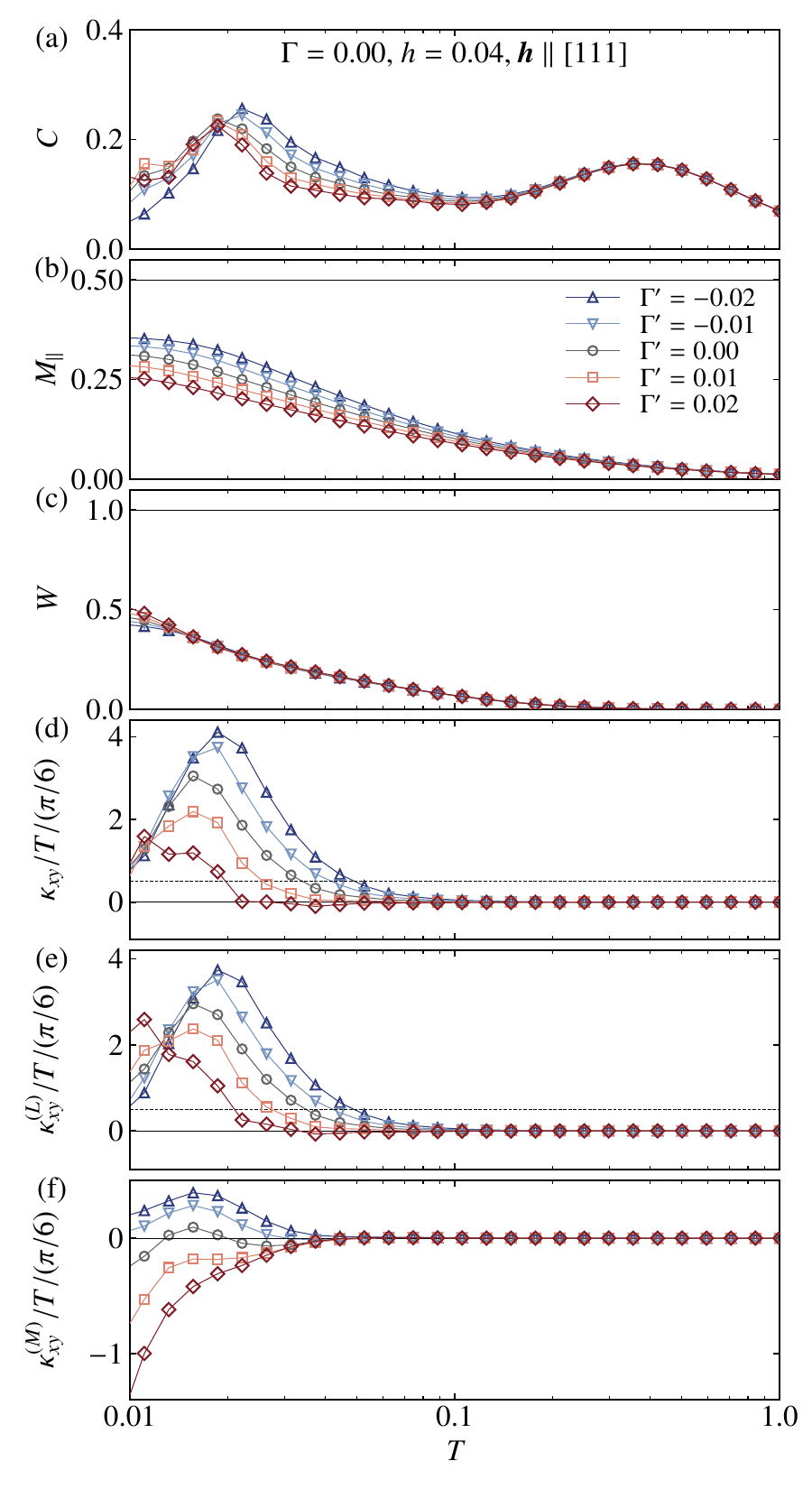}
  \end{center}
  \caption{Temperature dependence of (a) the specific heat (b) the magnetization, (c) the flux density, 
   (d) the total thermal Hall conductivity $\kappa_{xy}/T$, and
   (e) three-spin and (f) two-spin contributions to $\kappa_{xy}/T$ of the extended Kitaev model with $\Gamma=0$ and varying $\Gamma^{\prime}$ under the $[111]$ magnetic field with $h=0.04$.
  }
  \label{fig:all_h0.04_Gp}
\end{figure}

First, we consider the effects of the $\Gamma'$ interaction. 
Figure~\ref{fig:all_h0.04_Gp} displays the representative results at $h=0.04$ for $-0.02 \leq \Gamma' \leq 0.02$. 
Here, we set $\Gamma = 0$ to isolate the influence of the $\Gamma'$ term. 
As shown in Fig.~\ref{fig:all_h0.04_Gp}(a), $\Gamma^{\prime}$ does not significantly change 
the overall temperature dependence of the specific heat, although the low-temperature peak
shifts slightly to higher (lower) temperatures for negative (positive) $\Gamma'$. 
We also observe that the magnetization increases (decreases) monotonically with negative (positive) $\Gamma'$, as shown in Fig.~\ref{fig:all_h0.04_Gp}(b).
Compared with the results in Figs.~\ref{fig:CMF_pure}(a) and \ref{fig:CMF_pure}(b), this behavior can be interpreted by considering $\Gamma^{\prime}$ as an effective magnetic field, whose strength increases (decreases) for a negative (positive)  $\Gamma^{\prime}$. 
This interpretation appears to be consistent with the perturbation theory, which predicts that the Majorana band gap is proportional to $-\Gamma^{\prime}/|K|$~\cite{TakikawaF2020}. 
In contrast, as shown in Fig.~\ref{fig:all_h0.04_Gp}(c), the flux density does not show considerable changes as a function of $\Gamma^{\prime}$, suggesting limitations of the effective field picture.

Figures~\ref{fig:all_h0.04_Gp}(d)--\ref{fig:all_h0.04_Gp}(f) show the temperature dependence of total $\kappa_{xy}/T$ and its decomposition into the three-spin and two-spin components for the same values of $\Gamma'$. 
We find that $\kappa_{xy}/T$ increases (decreases) and its peak shifts to higher (lower) temperatures as $\Gamma'$ decreases (increases). 
This trend aligns with the effective field picture discussed above. 
However, the influence of $\Gamma'$ is more pronounced than the bulk physical quantities shown above. 
Indeed, we note that $\kappa_{xy}/T$ even changes sign at $\Gamma^{\prime} = 0.02$ for $h\simeq0.04$, although its absolute value is small. 
A similar trend is evident in the three-spin and two-spin contributions in Figs.~\ref{fig:all_h0.04_Gp}(e) and \ref{fig:all_h0.04_Gp}(f). 
The influence of $\Gamma^{\prime}$ is also significant in these components; in particular, a competition arises between the positive $\kappa_{xy}^{(L)}$ and the negative $\kappa_{xy}^{(M)}$ for $\Gamma^{\prime}\geq0.01$, resulting in a sign reversal in the total $\kappa_{xy}/T$ for large $\Gamma^{\prime}$. 

We observe similar overall trends for other values of the magnetic field. 
A detailed summary of the $\kappa_{xy}/T$ behavior on the $h$-$T$ plane will be provided in Sec.~\ref{sec:summary_of_temp_field}.

\subsubsection{Symmetric off-diagonal interaction $\Gamma$}\label{subsec:Gamma}
\begin{figure}[tbh]
  \begin{center}
    \includegraphics[width=\linewidth]{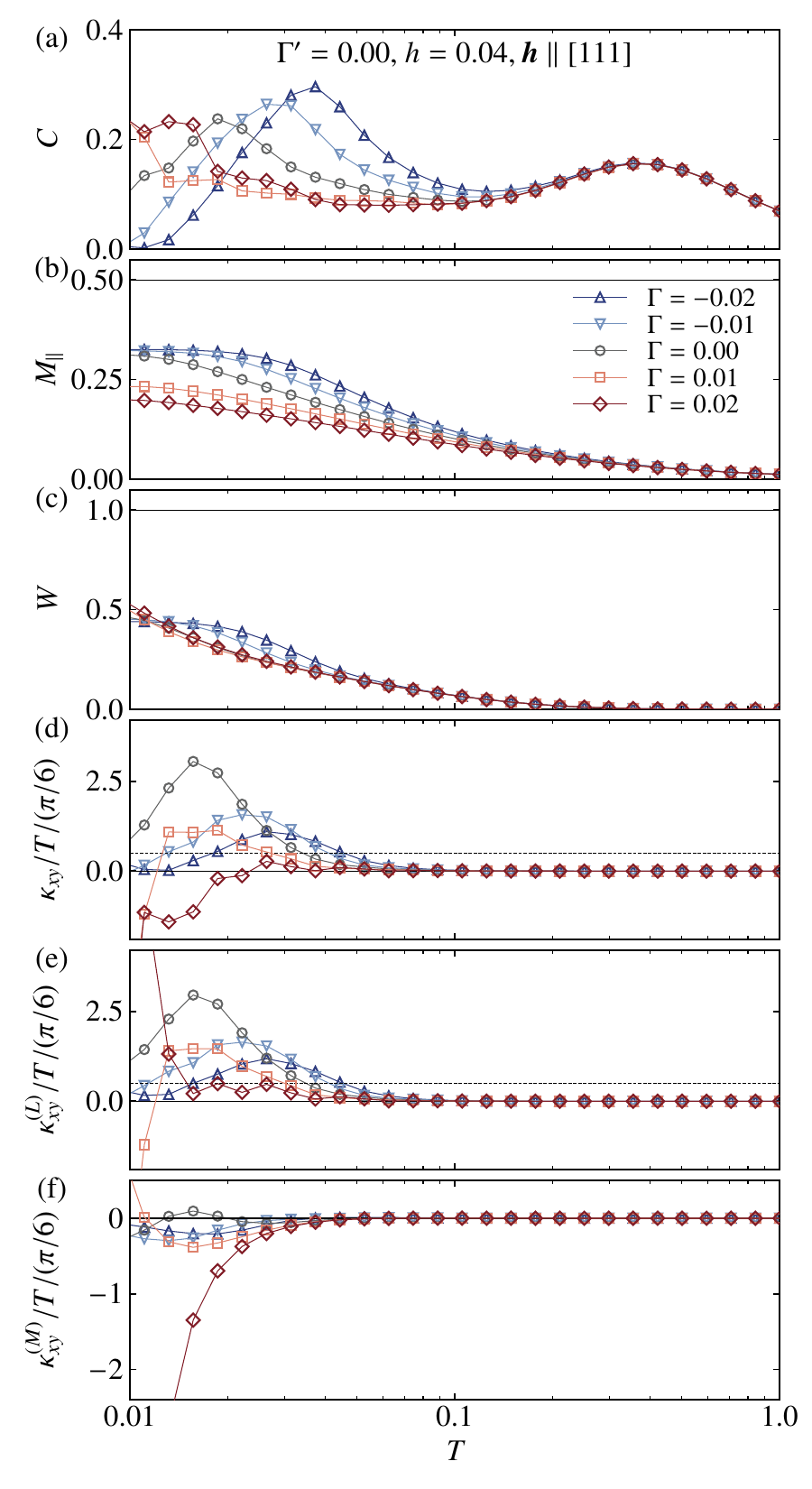}
  \end{center}
  \caption{
 Corresponding plots to Fig.~\ref{fig:all_h0.04_Gp} for the extended Kitaev model with $\Gamma^{\prime}=0$, $h=0.04$, and varying $\Gamma$. 
  }
  \label{fig:all_h0.04_G}
\end{figure}

\begin{figure}[tbh]
  \begin{center}
    \includegraphics[width=\linewidth]{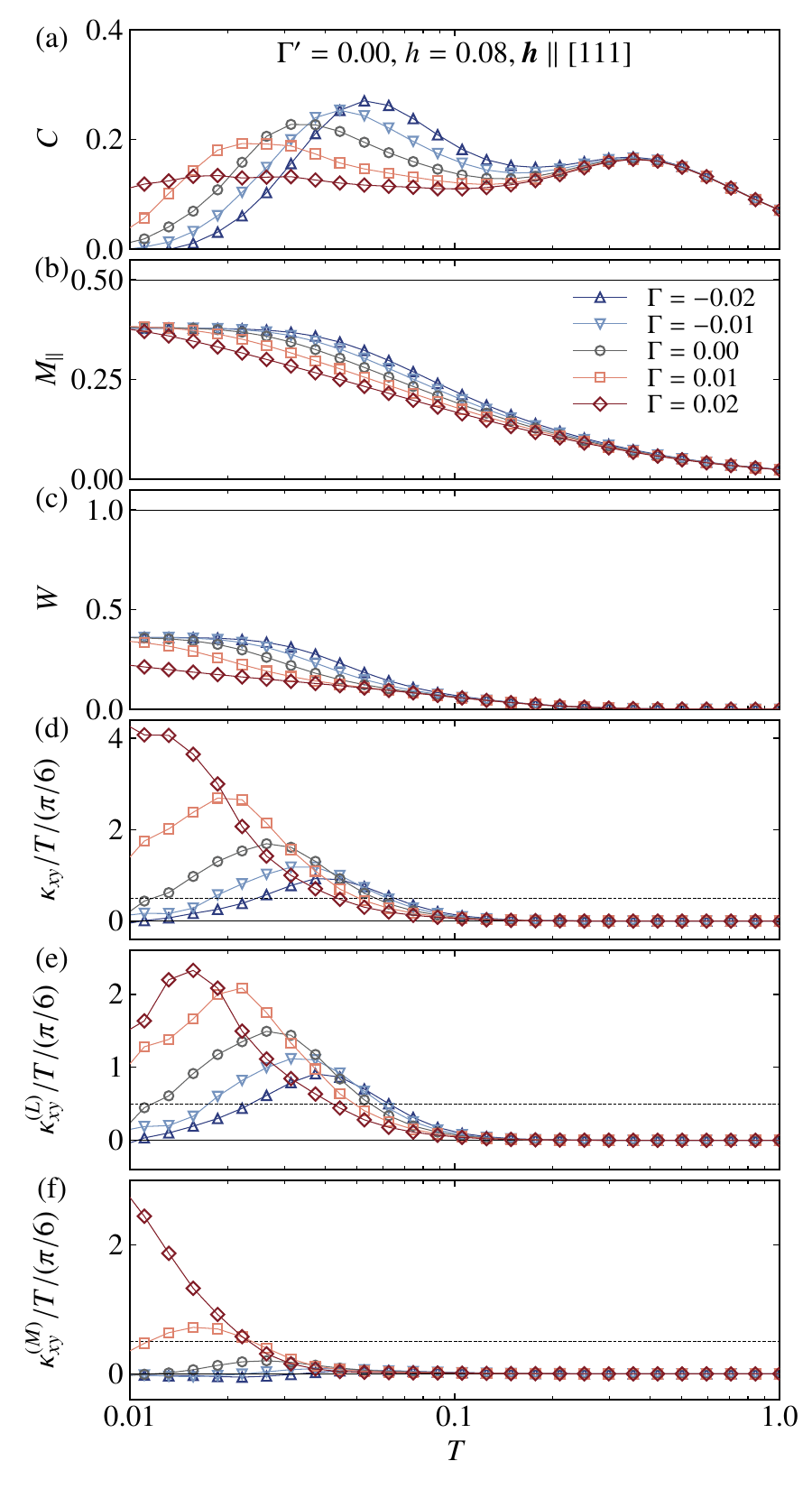}
  \end{center}
  \caption{
   Corresponding plots to Fig.~\ref{fig:all_h0.04_Gp} for the extended Kitaev model with $\Gamma^{\prime}=0$, $h=0.08$, and varying $\Gamma$. 
  }
  \label{fig:all_h0.08_G}
\end{figure}

Next, we investigate the effects of the $\Gamma$ interaction. 
Here, we consider two representative magnetic fields, $h=0.04$ (Fig.~\ref{fig:all_h0.04_G}) and $h=0.08$ (Fig.~\ref{fig:all_h0.08_G}).
Figures~\ref{fig:all_h0.04_G}(a)--\ref{fig:all_h0.04_G}(c) and \ref{fig:all_h0.08_G}(a)--\ref{fig:all_h0.08_G}(c) show the temperature dependence of the specific heat, the magnetization, and the flux density for $-0.02\leq\Gamma\leq0.02$. 
Although the magnitudes of $\Gamma$ are relatively small, we find that the low-temperature peak of the specific heat undergoes significant changes: 
it shifts to higher (lower) temperatures and its height increases (decreases) for negative (positive) $\Gamma$.
At $h=0.08$, the peak nearby disappears, and the specific heat exhibits less temperature dependence at low temperatures at $\Gamma=0.02$. 
Similarly, the magnetization increases (decreases) for negative (positive) $\Gamma$. 
We also observe $\Gamma$-sign-dependent changes in the flux density around the temperature region $T\sim0.03$, where the specific heat exhibits the low-temperature peak, although the effect is less pronounced. 


Figures~\ref{fig:all_h0.04_G}(d)--\ref{fig:all_h0.04_G}(f) and \ref{fig:all_h0.08_G}(d)--\ref{fig:all_h0.08_G}(f) present the results for $\kappa_{xy}/T$ and its decomposition. 
We find that the $\Gamma$ term also has a significant impact on the thermal Hall conductivity. 
A negative $\Gamma$ suppresses $\kappa_{xy}/T$ for both $h = 0.04$ and $h=0.08$, whereas a positive $\Gamma$ affects $\kappa_{xy}/T$ differently depending on the value of $h$.
For $h=0.04$, a positive $\Gamma$ strongly suppresses $\kappa_{xy}/T$, even driving it negative at $\Gamma = 0.02$, as shown in Fig.~\ref{fig:all_h0.04_G}(d). In contrast, at $h=0.08$, a positive $\Gamma$ enhances $\kappa_{xy}/T$, as shown in Fig.~\ref{fig:all_h0.08_G}(d). 
It is worth noting that these changes have not been predicted by the perturbation theory, where contributions from $\Gamma$ appear only at third order in terms of $h$~\cite{YamadaF2021}. 
The origin of the qualitative difference in the response to positive $\Gamma$ can be understood from the behavior of the three-spin and two-spin contributions, plotted in Figs.~\ref{fig:all_h0.04_G}
(e), \ref{fig:all_h0.04_G}(f), \ref{fig:all_h0.08_G}(e), and \ref{fig:all_h0.08_G}(f). 
For $h=0.08$, both contributions vary smoothly with $\Gamma$. 
However, for $h=0.04$, they exhibit abrupt changes at low temperatures for positive $\Gamma$. 
These abrupt changes may be attributed to a phase transition from the Kitaev QSL to another phase, such as a spin nematic state, as suggested by infinite tensor network calculations at zero temperature~\cite{Lee_NCom2020}.

\subsection{Summary of $h$-$T$ dependence}
\label{sec:summary_of_temp_field}
\begin{figure*}[htb]
  \begin{center}
    \includegraphics[width=\linewidth]{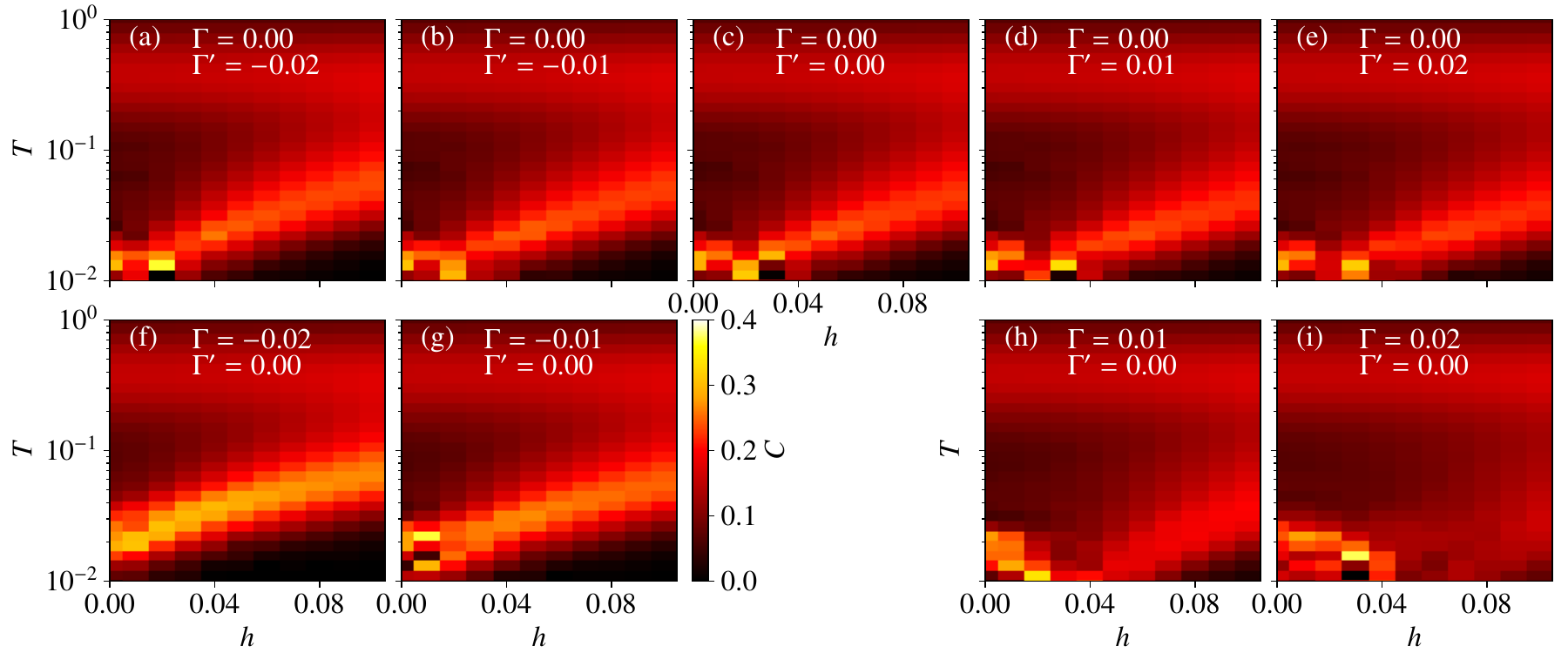}
  \end{center}
  \caption{
Color maps of the specific heat for the pure Kitaev model on the field-temperature plane with (a)-(e) $\Gamma=0$ and varying $\Gamma'$ and  (f)-(i) varying $\Gamma$ and $\Gamma'=0$.}
  \label{fig:color_map_C}
\end{figure*}

\begin{figure*}[htb]
  \begin{center}
    \includegraphics[width=\linewidth]{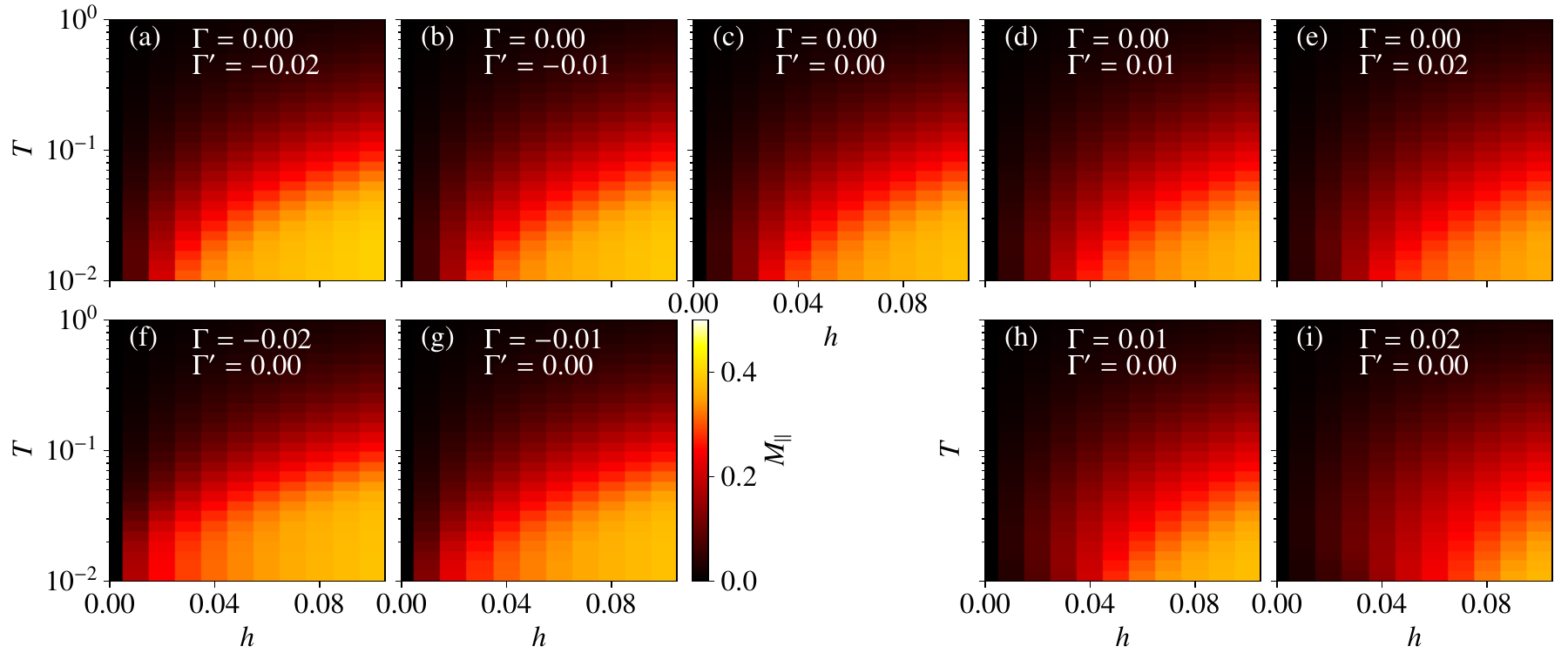}
  \end{center}
  \caption{
Color maps of the magnetization, corresponding to Fig.~\ref{fig:color_map_C}.} 
  \label{fig:color_map_mag}
\end{figure*}

\begin{figure*}[htb]
  \begin{center}
    \includegraphics[width=\linewidth]{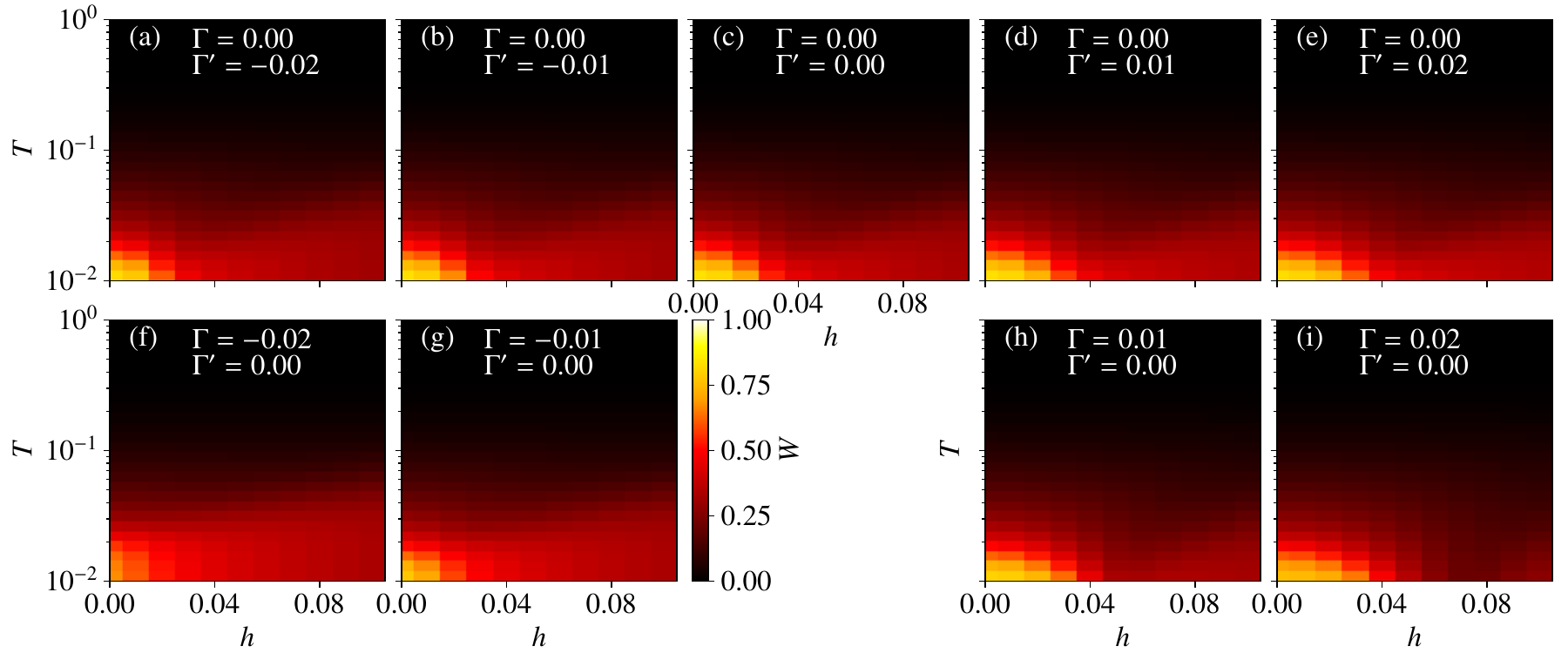}
  \end{center}
  \caption{
Color maps of the flux density, corresponding to Fig.~\ref{fig:color_map_C}.} 
  \label{fig:color_map_flux}
\end{figure*}

\begin{figure*}[htb]
  \begin{center}
    \includegraphics[width=\linewidth]{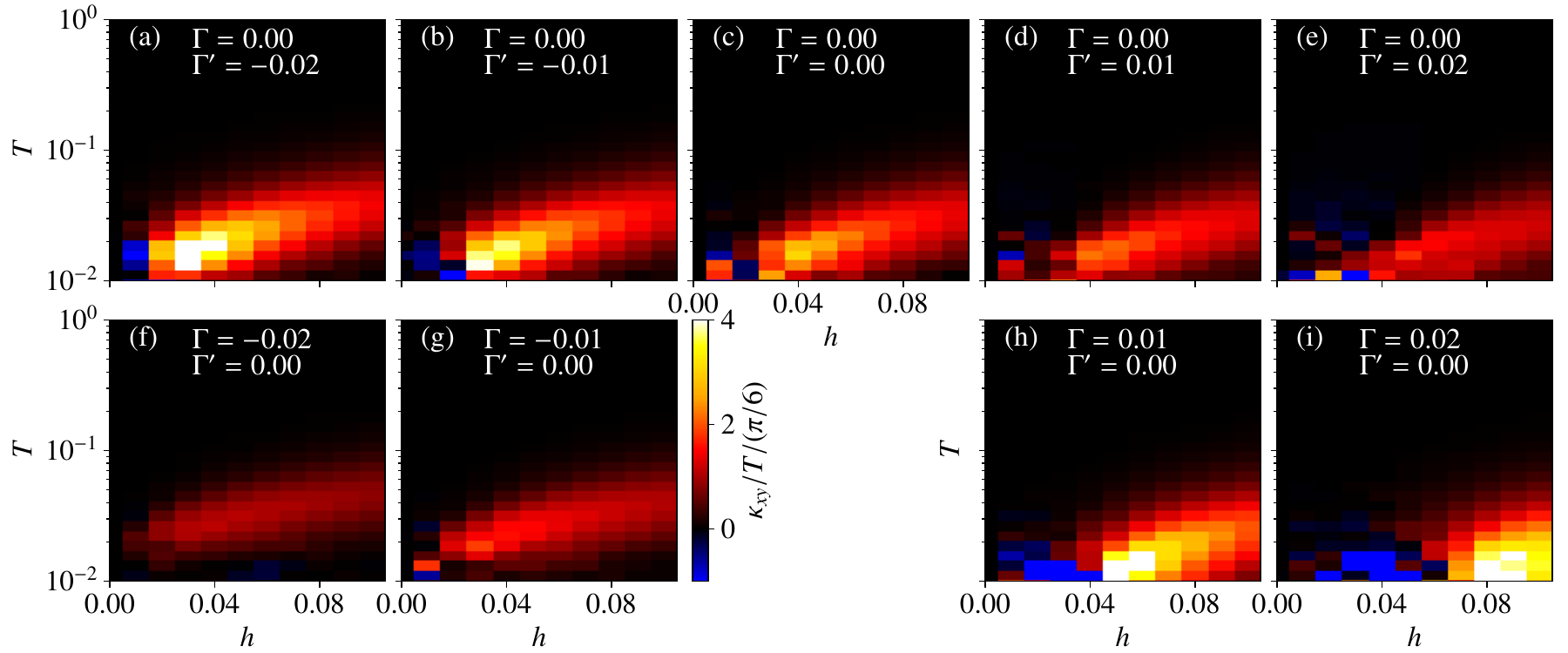}
  \end{center}
  \caption{
Color maps of $\kappa_{xy}/T$, corresponding to Fig.~\ref{fig:color_map_C}.} 
  \label{fig:color_map_all}
\end{figure*}

Summarizing results in the previous subsections, we plot the temperature and field dependence of the specific heat, magnetization, flux density, and $\kappa_{xy}/T$ in Figs.~\ref{fig:color_map_C}, \ref{fig:color_map_mag}, \ref{fig:color_map_flux}, and \ref{fig:color_map_all}, respectively, for various values of $\Gamma$ and $\Gamma'$. 
In each figure, the central panel corresponds to the result for the pure Kitaev model with $K=-1$ and $\Gamma=\Gamma'=0$. 
The upper panels display the results for varying $\Gamma'$ with $\Gamma=0$, and the lower panels show the results for varying $\Gamma$ with $\Gamma'=0$. 

From the specific heat shown in Fig.~\ref{fig:color_map_C}, we observe that the low-temperature peak exhibits a pronounced dependence on the magnetic field, whereas the high-temperature peak remains almost unchanged. For example, in the pure Kitaev model, as shown in the central panel, increasing the magnetic field initially reduces the peak temperature up to $h \sim 0.02$, beyond which it increases monotonically. This behavior is closely associated with the quantum critical point at $h_c \simeq 0.024$, as discussed in Sec.~\ref{subsec:pureKitaev_h111}. 
The low-temperature, weak-field regime corresponds to a quantum spin liquid, the low-temperature, strong-field regime corresponds to a polarized ferromagnet, and the intermediate temperature range between the low- and high-temperature peaks is interpreted as a fractionalized paramagnet. Similar behavior has been reported in previous studies~\cite{YoshitakeNKM2020,Li2020,LiLXGQLS2024}.
Although more sophisticated studies are needed to obtain reliable results in the low-temperature, weak-field regime as discussed in the end of Sec.~\ref{subsec:pureKitaev_h111}, the quantum critical point exhibits slight shifts upon varying the values of $\Gamma$ and $\Gamma'$, and in both cases, $h_c$ appears to increase when these parameters take positive values. 

These observations are consistent with the behavior of both magnetization and flux density. 
As shown in Fig.~\ref{fig:color_map_mag}, the magnetization begins to develop below the temperature corresponding to the low-temperature specific heat peak. 
This trend supports the presence of a polarized ferromagnetic state in this low-temperature, strong-field region ($h \gtrsim h_c$). 
Conversely, as shown in Fig.~\ref{fig:color_map_flux}, the flux density develops in the low-temperature, weak-field region ($h \lesssim h_c$), which aligns with the interpretation of this region as a quantum spin liquid.

Figure~\ref{fig:color_map_all} summarizes the behavior of $\kappa_{xy}/T$. 
For all values of $\Gamma$ and $\Gamma'$, we observe bright regions at low temperature that extend to higher temperatures under stronger magnetic fields. 
These regions correspond to the peaks in $\kappa_{xy}/T$, which overshoot the half-integer quantized value. 
Introducing a weak $\Gamma^{\prime}$ interaction significantly affects these peaks: 
negative $\Gamma'$ enhances them, while positive $\Gamma'$ suppresses them, as observed in Sec.~\ref{subsec:Gamma_prime}. 
The negative values of $\kappa_{xy}/T$ observed at weak fields and low temperatures are likely due to limited numerical accuracy, stemming from insufficient bond dimension $D$. 
In contrast, the introduction of $\Gamma$ produces the opposite trend:
positive $\Gamma$ enhances the peaks, whereas negative $\Gamma$ suppresses them, as discussed in Sec.~\ref{subsec:Gamma}.
In addition, positive $\Gamma$ makes $\kappa_{xy}/T$ negative in the weak-field, low-temperature regime, and this region appears to expand with increasing $\Gamma$. 
This behavior may signal a possible quantum phase transition, as also discussed in Sec.~\ref{subsec:Gamma}. 

As discussed in Sec.~\ref{sec:pureKitaev}, the pronounced overshooting behavior of $\kappa_{xy}/T$ is attributed to contributions from topological Majorana fermions. 
This interpretation is supported by analyses of the decomposition into three- and two-spin contributions, as well as the dependence on the field direction. 
The topological Majorana fermion picture, even in the polarized regime well beyond critical field, is also robust in the presence of $\Gamma$ and $\Gamma'$ interactions. 
For the extended Kitaev model with $\Gamma$ and $\Gamma'$ interactions, $\kappa_{xy}/T$ was studied using linear spin-wave theory based on the conventional magnon picture~\cite{ChernZK2021,ZhangCK2021}. 
It was shown that the contribution from topological magnons can alter the sign of $\kappa_{xy}/T$, depending not only on the field directions but also on the interaction parameters. 
Considering that magnon-magnon interactions beyond the linear spin-wave theory can also alter the sign of $\kappa_{xy}/T$ in the pure Kitaev model~\cite{Koyama2024}, further studies are necessary to elucidate the role of topological magnons in the extended Kitaev models.

Finally, let us remark on the correlation between the specific heat $C$ and $\kappa_{xy}/T$. 
As discussed in Sec.~\ref{subsec:pureKitaev_h111} for the pure Kitaev model, there is a clear correspondence between the low-temperature peak in $C$ and the overshooting peak in $\kappa_{xy}/T$. 
This persists even when the $\Gamma$ and $\Gamma'$ interactions are introduced, as shown in Figs.~\ref{fig:color_map_C} and \ref{fig:color_map_all}. 
This observation suggests that the enhancement of $\kappa_{xy}/T$ is closely related to the entropy release occurring in this temperature regime, which predominantly arises from the fluxes under thermal fractionalization in the pure Kitaev model~\cite{NasuUM2015}. 
Therefore, the correspondence between $C$ and $\kappa_{xy}/T$ highlights their shared origin in fractional excitations, offering a compelling experimental test for the origin of thermal Hall response. 
When such a correspondence is observed, it strongly indicates that the thermal Hall effect is primarily governed by spin degrees of freedom, rather than by other contributions such as phonons. 
However, given that the experimental data are currently available in the limited range of temperatures and fields, further systematic studies are highly anticipated. 

\subsection{Classical limit}
\label{sec:classicalMC}

In this subsection, we show the results for the classical counterpart of Eq.~\eqref{eq:model-Hamiltonian} and compare them with the quantum results discussed above.
The following numerical data are obtained using the classical Monte Carlo simulation method described in Sec.~\ref{subsec:Classical Monte Carlo simulation}. 

\begin{figure}[tbh] 
\begin{center} 
\includegraphics[width=\linewidth]{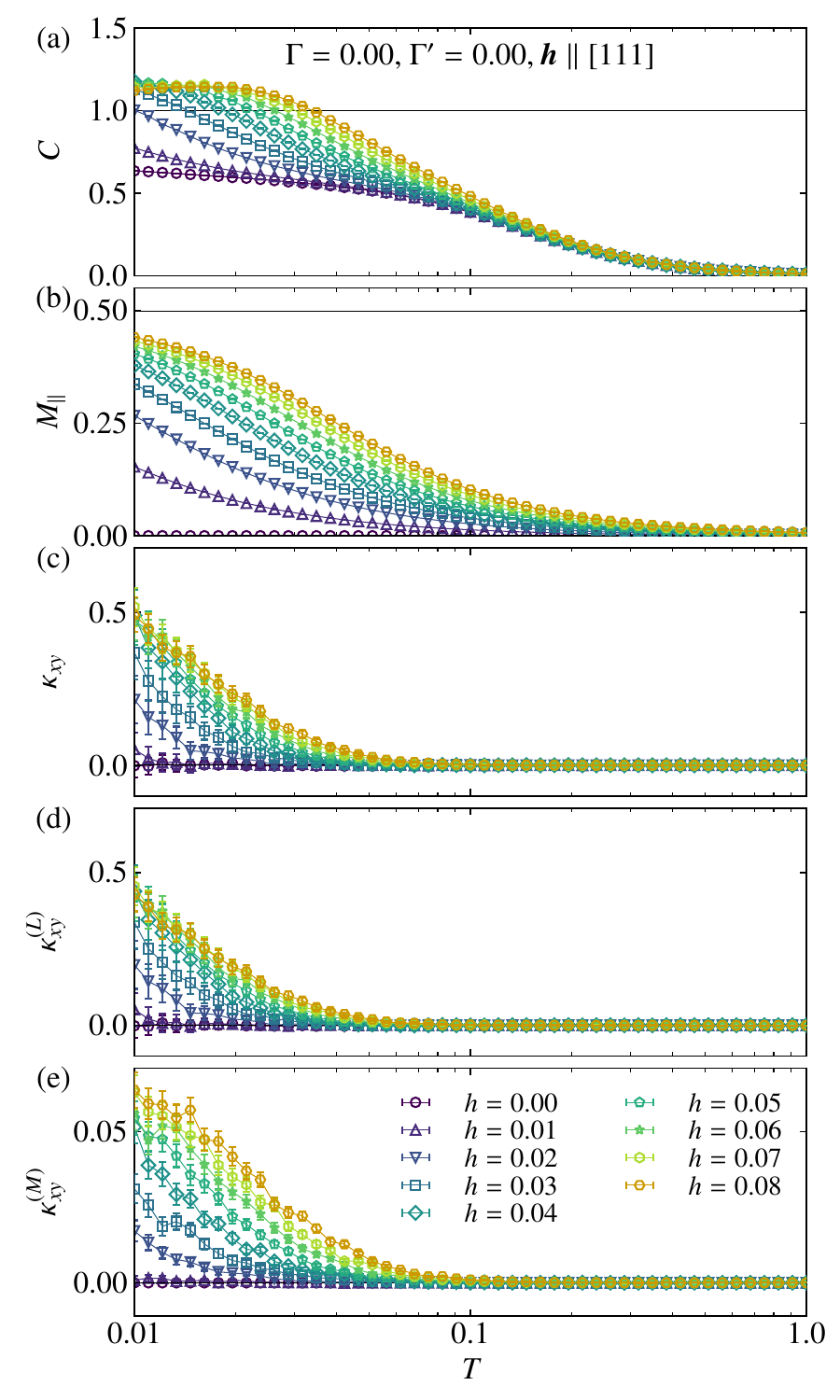}
\vspace{-0.5cm} 
\caption{Temperature dependence of (a) the specific heat, (b) the magnetization, (c) the thermal Hall conductivity $\kappa_{xy}$, and (d) three-spin and (e) two-spin contributions to $\kappa_{xy}$ of the classical Kitaev model for several magnetic fields along the $[111]$ direction.}
\label{fig_classical_hdep}
\end{center}
\end{figure}

First, we focus on the pure Kitaev model with $\Gamma=\Gamma'=0$ in the magnetic fields applied along the $[111]$ direction.
Figure~\ref{fig_classical_hdep}(a) shows the temperature dependence of the specific heat of the classical Kitaev model at several magnetic field strengths. 
In the absence of the magnetic field, the specific heat increases monotonically with decreasing temperature, and does not exhibit the double-peak structure observed in the quantum case. 
Previous studies have shown that it approaches $3/4$ in the zero temperature limit, which is significantly reduced from the expected value of $1$ for conventional classical spin models~\cite{Sela2014,Suzuki2018_2}.
This value reduction originates from the presence of zero modes and the associated quartic-order spin fluctuations, which are intrinsic to the pure Kitaev model without a magnetic field.
Once the magnetic field is introduced, these zero modes are lifted, and the specific heat approaches $1$ in the zero temperature limit. 
We find that, during the additional rise at low temperatures, the specific heat exhibits an inflection point at the temperature corresponding to the energy scale of $h$. 
As shown in Fig.~\ref{fig_classical_hdep}(b), the magnetization also exhibits an inflection point around the same temperature.

Figure~\ref{fig_classical_hdep}(c) shows the thermal Hall conductivity $\kappa_{xy}$ as a function of temperature.
In the classical system, since $\kappa_{xy}/T$ does not exhibit quantization and tends to diverge in the low-temperature limit, we discuss $\kappa_{xy}$ rather than $\kappa_{xy}/T$.
We observe that $\kappa_{xy}$ is zero at zero field, and becomes nonzero and positive by introducing the magnetic field.  
This behavior is qualitatively consistent with the quantum results shown in Fig.~\ref{fig:CMF_pure}(d).
Moreover, we find that $\kappa_{xy}$ begins to develop at much lower temperatures compared to the onset of the magnetization.
These distinctly different temperature scales are also observed in the quantum system (see Fig.~\ref{fig:CMF_pure}). 
However, in the classical case, $\kappa_{xy}$ increases monotonically with decreasing temperature, in contrast to the quantum result, which exhibits the overshooting behavior with a peak around the low-temperature peak of the specific heat. 

Figures~\ref{fig_classical_hdep}(d) and \ref{fig_classical_hdep}(e) present the three-spin and two-spin contributions to the thermal Hall conductivity, respectively.
The two-spin contribution $\kappa_{xy}^{(M)}$ is much smaller than the three-spin 
one $\kappa_{xy}^{(L)}$, indicating that the thermal Hall effect is dominated by the three-spin terms of the thermal current at the edges.
This trend is consistent with the behavior observed in the quantum system. 
However, $\kappa_{xy}^{(M)}$ in the classical approximation is always positive and does not reproduce the small negative values of $\kappa_{xy}^{(M)}/T$ in weak magnetic fields in Fig.~\ref{fig:CMF_pure}(f). 
This discrepancy suggests that the negative $\kappa_{xy}^{(M)}/T$ can be attributed to quantum effects.

\begin{figure}[tbh] 
\begin{center} 
\includegraphics[width=\linewidth]{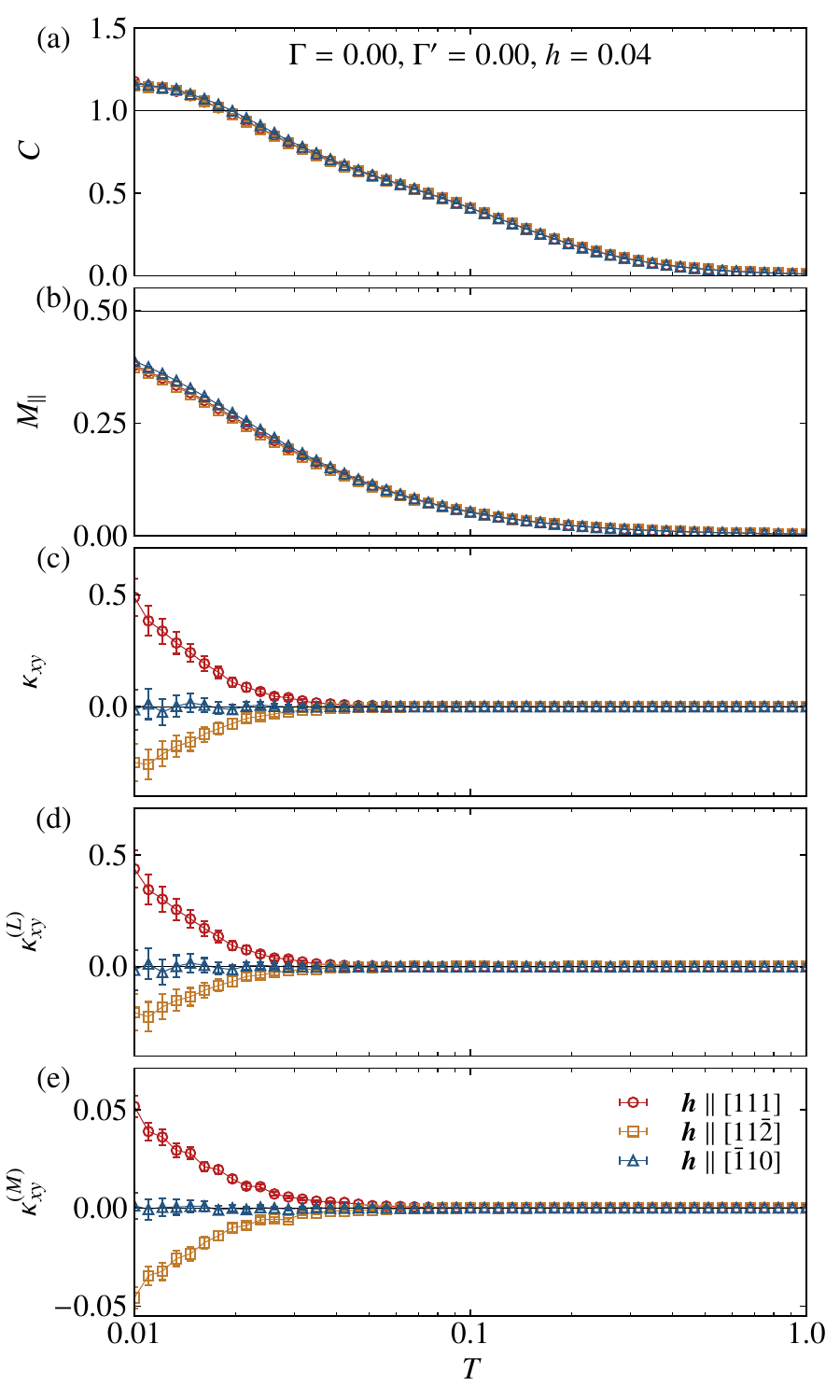}
\vspace{-0.5cm} 
\caption{Corresponding plots to Fig.~\ref{fig_classical_hdep} under the magnetic field parallel to $[111]$, $[11\bar{2}]$, and $[\bar{1}10]$ with $h=0.04$.
}
\label{fig_classical_adep004}
\end{center}
\end{figure}
  
Next, we discuss the field-direction dependence of physical quantities in the pure Kitaev model. 
Following the quantum case in Sec.~\ref{subsec:field_direction_dep}, we consider three mutually-perpendicular field directions: $[11\bar{2}]$, $[\bar{1}10]$, and $[111]$.
Figures~\ref{fig_classical_adep004}(a) and \ref{fig_classical_adep004}(b) show the temperature dependence of the specific heat and magnetization, respectively.
These results indicate that the field direction has little effect on either quantity, similar to the quantum system (see Fig.~\ref{fig:CMF_ab}). 
In contrast to these bulk physical quantities, the thermal Hall conductivity strongly depends on the direction of the applied magnetic field.
As shown in Fig.~\ref{fig_classical_adep004}(c), $\kappa_{xy}$ is positive and increases with decreasing temperature for $\vec{h}\parallel [111]$, whereas it is negative and decreases for
$\vec{h}\parallel [11\bar{2}]$; 
it vanishes for the $\vec{h}\parallel[\bar{1}10]$, consistent with the symmetry argument in the end of Sec.~\ref{subsec:field_direction_dep}.
Moreover, we observe that the absolute value for $\vec{h}\parallel [111]$ is larger than that for $\vec{h}\parallel [11\bar{2}]$.
While these trends are consistent with those observed in the quantum system, 
the overshooting behavior with a peak structure in Fig.~\ref{fig:CMF_ab}(d) is not reproduced in the classical case.
Furthermore, although the three-spin contribution $\kappa_{xy}^{(L)}$, shown in Fig.~\ref{fig_classical_adep004}(d), follows these trends, the two-spin contribution $\kappa_{xy}^{(M)}$, shown in Fig.~\ref{fig_classical_adep004}(e), remains small and fails to reproduce the quantum behavior presented in Fig.~\ref{fig:CMF_ab}(f). 
This result supports the significant role of quantum fluctuations in the two-spin contribution to the thermal Hall effect.

\begin{figure}[tbh] 
\begin{center} 
\includegraphics[width=\linewidth]{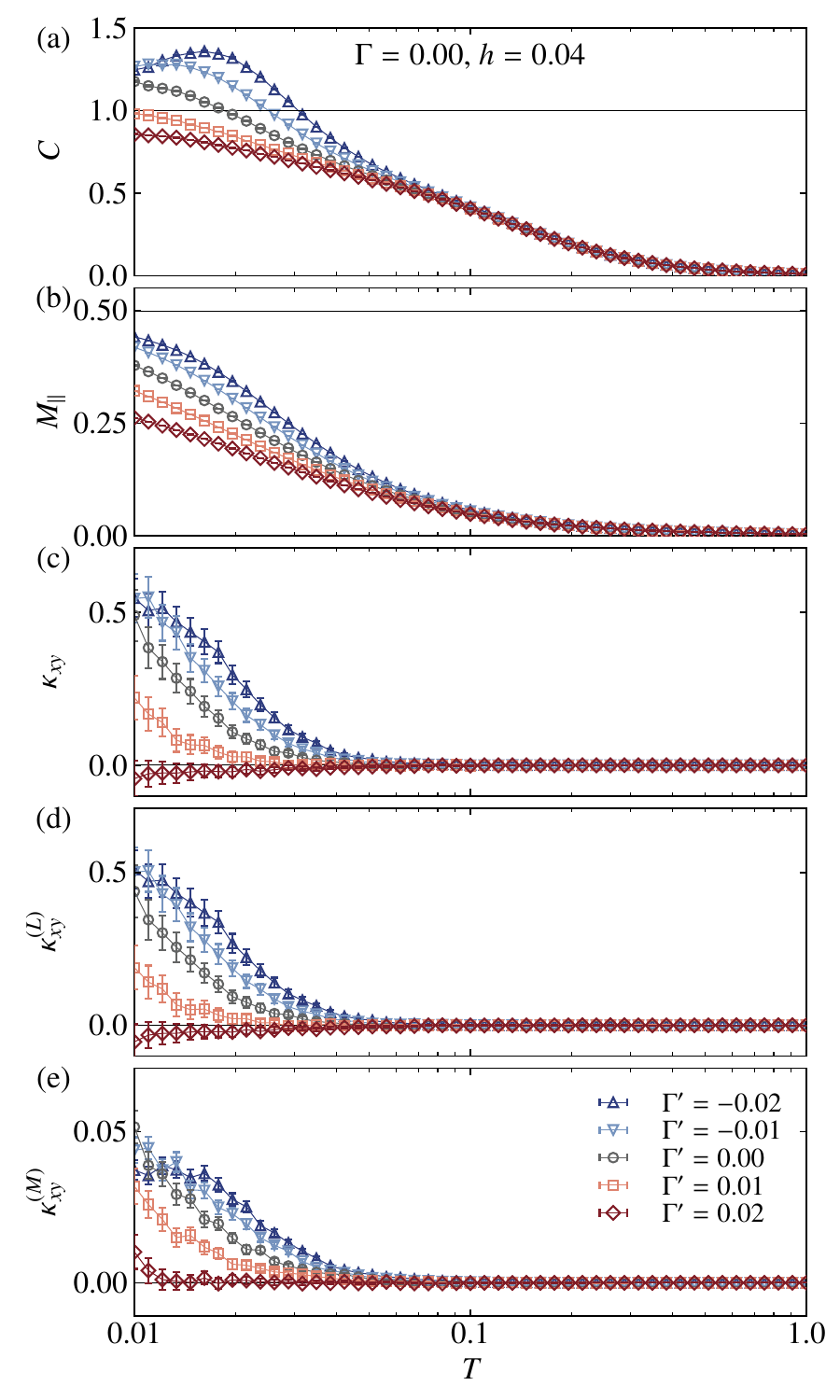}
\vspace{-0.5cm} 
\caption{Corresponding plots to Fig.~\ref{fig_classical_hdep} for the classical extended Kitaev model with $\Gamma=0$ and varying $\Gamma'$ under the $[111]$ magnetic field with $h=0.04$.
}
\label{fig_classical_gpdep004}
\end{center}
\end{figure}

Now, we turn to the effects of the $\Gamma'$ and $\Gamma$ interactions. 
Figure~\ref{fig_classical_gpdep004} shows the temperature dependence of physical quantities for several values of $\Gamma'$, with $\Gamma$ set to zero. 
We observe that a positive (negative) $\Gamma'$ suppresses (enhances) both the specific heat and magnetization.
Moreover, a negative $\Gamma'$ induces a peak in the specific heat, which shifts to higher temperatures with decreasing $\Gamma'$.
These trends are also observed in the quantum system, as shown in Figs.~\ref{fig:all_h0.04_Gp}(a) and \ref{fig:all_h0.04_Gp}(b), although the double-peak structure in the specific heat is not reproduced in the classical results.

Regarding thermal transport, we find that a negative (positive) $\Gamma'$ enhances (suppresses) the thermal Hall conductivity, as shown in Fig.~\ref{fig_classical_gpdep004}(c).
We note that $\kappa_{xy}$ becomes negative at low temperatures for $\Gamma'=0.02$.
We also find that the three-spin term provides the dominant contribution to the thermal Hall conductivity, as shown in Figs.~\ref{fig_classical_gpdep004}(d) and \ref{fig_classical_gpdep004}(e).
Overall, these trends are consistent with those observed in the quantum systems.
It is also worth noting that the classical results at low temperatures qualitatively reproduce the quantum behavior in the intermediate temperature range ($T\sim 0.05$), as shown in Figs.~\ref{fig:all_h0.04_Gp}(d)--\ref{fig:all_h0.04_Gp}(f). 
This indicates that $\kappa_{xy}$ in the quantum systems at intermediate temperatures, where quantum fluctuations are less important, can be effectively described by its classical counterpart.

\begin{figure}[t] 
\begin{center} 
\includegraphics[width=\linewidth]{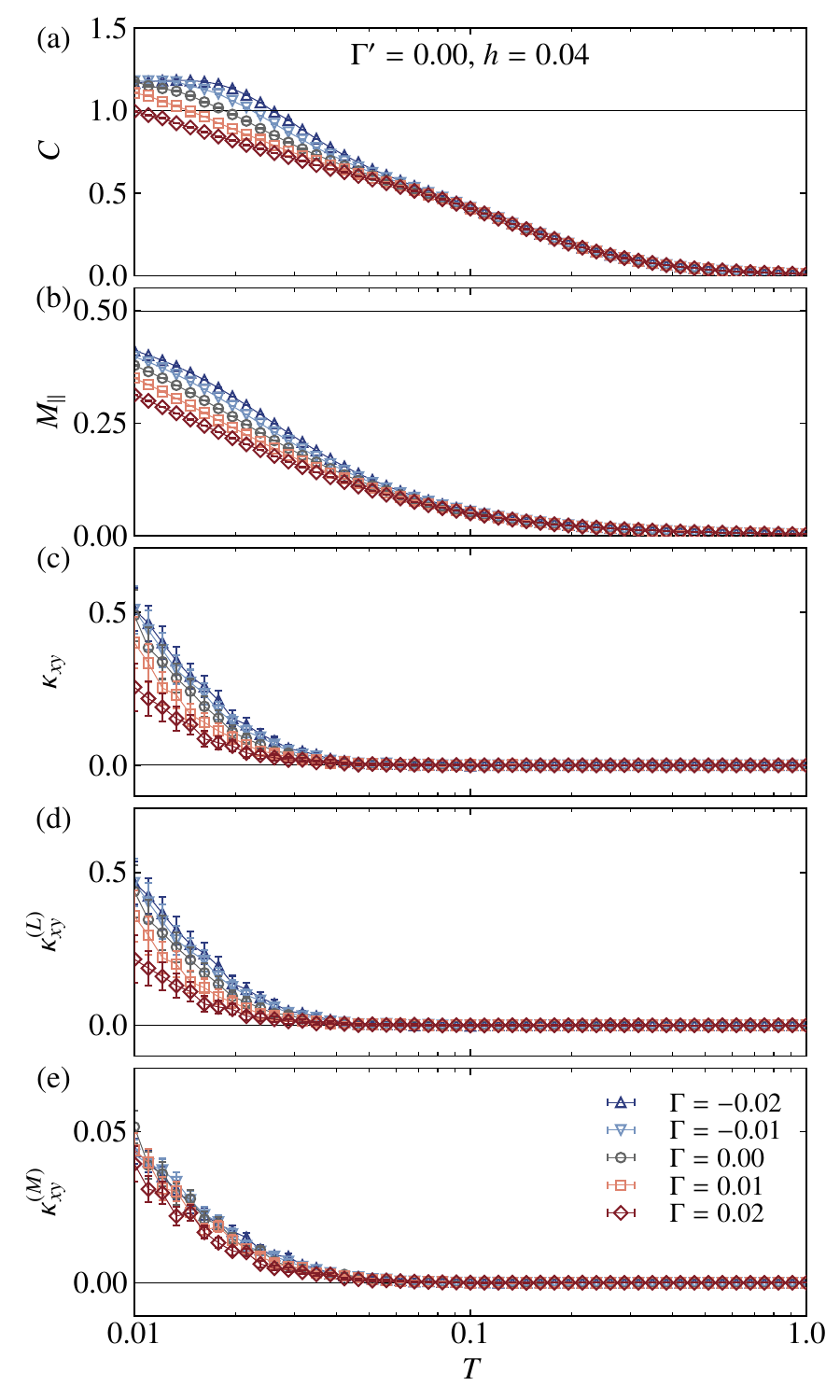}
\vspace{-0.5cm} 
\caption{
Corresponding plots to Fig.~\ref{fig_classical_hdep} for the classica Kitaev model with $\Gamma'=0$ and varying $\Gamma$ under the $[111]$ magnetic field with $h=0.04$.}
\label{fig_classical_gdep004}
\end{center}
\end{figure}

Figure~\ref{fig_classical_gdep004} shows the results for several values of $\Gamma$, setting $\Gamma'=0$.
For negative $\Gamma$, the low-$T$ specific heat increases with the absolute value of $\Gamma$, as shown in Fig.~\ref{fig_classical_gdep004}(a), while introducing a positive $\Gamma$ leads to a suppression of the specific heat.
These trends are again consistent with the quantum results, except for the absence of the double-peak structure.
In addition, as shown in Fig.~\ref{fig_classical_gdep004}(b), the magnetization increases with decreasing $\Gamma$, which is also observed in the quantum 
system, as shown in Fig.~\ref{fig:all_h0.04_G}(b).

Nonetheless, as shown in Figs.~\ref{fig_classical_gdep004}(c)-\ref{fig_classical_gdep004}(e), thermal Hall tranport in the presence of $\Gamma$ exhibits behavior distinct from that of the quantum system in Figs.~\ref{fig:all_h0.04_G}(d)-\ref{fig:all_h0.04_G}(f).
A key feature of the quantum system is substantial changes of both $\kappa_{xy}^{(L)}$ and $\kappa_{xy}^{(M)}$, which nearly cancel each other out in the total thermal Hall conductivity $\kappa_{xy}$.
This feature is not reproduced in the classical system; $\kappa_{xy}^{(L)}$ changes with $\Gamma$, whereas $\kappa_{xy}^{(M)}$ is almost intact. 
This contrasting behavior suggests that the cancellation between $\kappa_{xy}^{(L)}$ and $\kappa_{xy}^{(M)}$ originates from quantum fluctuations.
However, similar to the case with $\Gamma'$, we again observe that the $\Gamma$ and temperature dependencies of $\kappa_{xy}$ and its decomposition in the intermediate temperature region at $T \sim0.05$, shown in Figs.~\ref{fig:all_h0.04_G}(d)--\ref{fig:all_h0.04_G}(f), are qualitatively reproduced by the low-temperature behavior in the classical system, as shown in Figs.~\ref{fig_classical_gdep004}(c)--\ref{fig_classical_gdep004}(e).

Finally, in comparing the effects of $\Gamma^{\prime}$ and $\Gamma$ terms, we find that both the specific heat and magnetization appear to be more sensitive to $\Gamma'$ than to $\Gamma$. 
This trend contrasts with that observed in the quantum system, as shown in Figs.~\ref{fig:all_h0.04_Gp} and \ref{fig:all_h0.04_G}.
The disparity suggests that the $\Gamma$ interaction induces more significant quantum effects compared to
the $\Gamma'$ interaction.
Indeed, previous studies have pointed out that the $\Gamma$ interaction does not destroy the Kitaev QSL state~\cite{Gohlke_PRB2018,catuneanu2018}, whereas the introduction of $\Gamma'$ tends to shrink the region of the QSL phase~\cite{gordon2019theory,Luo2022}.

\begin{figure*}[tbh]
  \begin{center}
    \includegraphics[width=\linewidth]{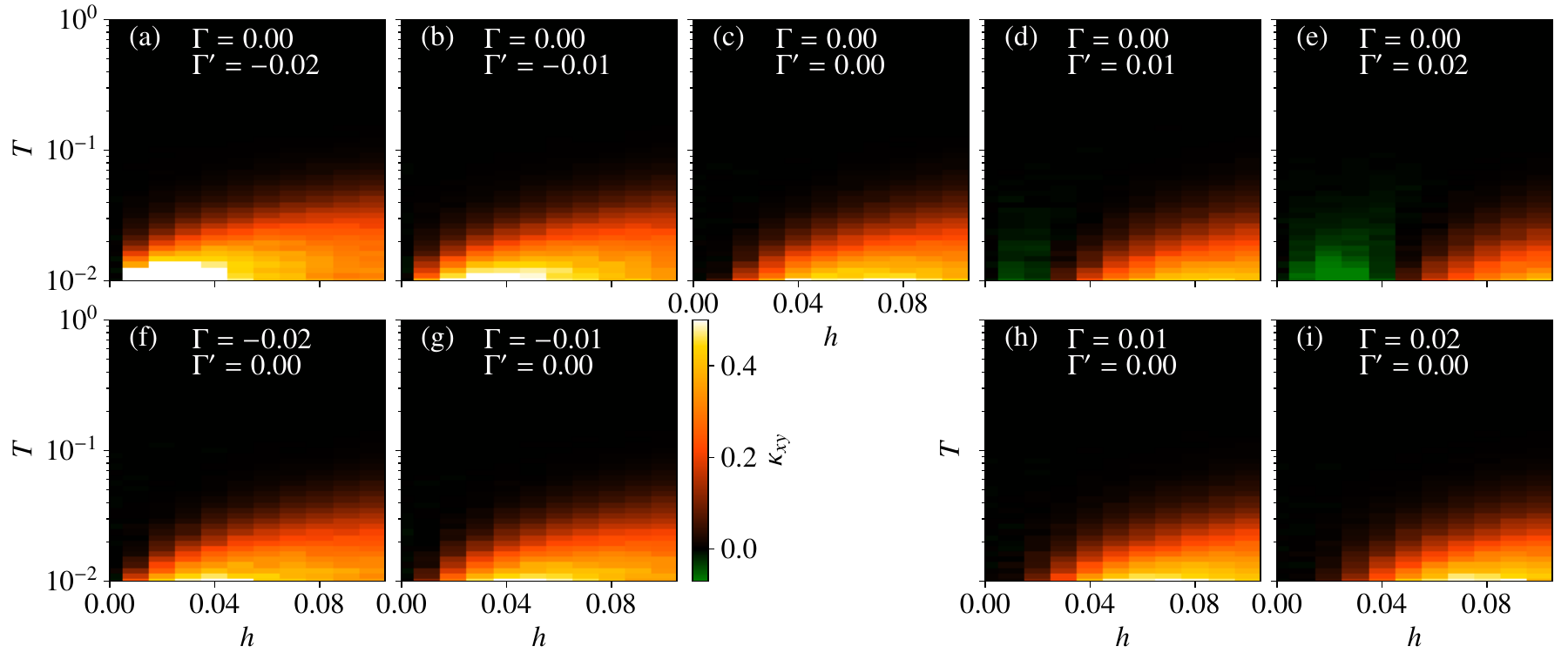}
  \end{center}
  \caption{
Color maps of $\kappa_{xy}$ for the classical model. The panel arrangement is common to Fig.~\ref{fig:color_map_all}.
  }
  \label{fig:cmap_classical}
\end{figure*}

Summarizing the classical results in this subsection, we plot the temperature and field dependence of the thermal Hall conductivity in Fig.~\ref{fig:cmap_classical} for various values of $\Gamma$ and $\Gamma'$.
Note that $\kappa_{xy}$ is shown for the classical case in Fig.~\ref{fig:cmap_classical}, while $\kappa_{xy}/T$ is presented for the quantum case in Fig.~\ref{fig:color_map_all}.
In the classical case, the $\Gamma'$ interaction leads to qualitatively similar behavior to the quantum case, except that a clear negative region appears in the weak-field, low-temperature regime for positive $\Gamma'$. 
In contrast, the $\Gamma$ interaction has little effect on the overall behavior of $\kappa_{xy}$.
This insensitivity to the $\Gamma$ interaction stands in stark contrast to the results for quantum systems.
These findings underscore the distinct nature of $\Gamma$ and $\Gamma'$ interactions: 
while the effects of $\Gamma'$ interaction can be captured within the classical framework, the influence of $\Gamma$ is inherently quantum, as discussed above. 

\section{Summary}
\label{sec:Summary}
In this study, we have conducted a systematic and comprehensive analysis of the thermal Hall conductivity $\kappa_{xy}$ in the extended Kitaev model, using the tensor network representation of the density matrix. 
This method enables highly accurate evaluation of the thermal Hall response beyond conventional perturbation theory. 
We have explored the temperature and magnetic field dependence across a broad range of additional non-Kitaev interactions, $\Gamma$ and $\Gamma'$. 

An important finding of this study is that $\kappa_{xy}/T$ significantly exceeds the half-integer quantized value at intermediate temperatures, even within the pure Kitaev model with $\Gamma=\Gamma'=0$. 
The enhancement arises predominantly from the three-spin contributions $\kappa_{xy}^{(L)}$, suggesting that it may be attributed to the contributions from Majorana fermions inherent in the Kitaev model. 
In addition, we demonstrated that $\kappa_{xy}/T$ changes its sign depending on the direction of the magnetic field. 
This behavior is consistent with the sign of the Chern number of Majorana fermion bands predicted by perturbation theory, thereby supporting the topological origin of the thermal Hall effect. 
Notably, this sign change as well as the overshooting behavior persists even under strong magnetic fields, indicating that topological Majorana contributions remain robust well into the high-field polarized regime, beyond the quantum critical point. 

It is worth emphasizing that the pronounced overshooting behavior of $\kappa_{xy}/T$ beyond the half-integer quantized value has not been observed in the previous numerical study of the Kitaev model~\cite{KumarT2023}. 
The discrepancy arises from the proper definitions of polarization and energy current, including their position dependence, which was not taken into account in the previous study. 

We have also demonstrated that the thermal Hall response is significantly affected by 
the $\Gamma$ and $\Gamma^{\prime}$ interactions, even when their magnitudes are two orders of magnitude smaller than the Kitaev interaction. 
In particular, for negative $\Gamma^{\prime}$ and positive $\Gamma$, $\kappa_{xy}/T$ is enhanced, though the underlying mechanisms differ substantially.
In the case of negative $\Gamma^{\prime}$, we find that the enhancement is primarily driven by the three-spin term $\kappa_{xy}^{(L)}$, consistent with the scenario in which the negative $\Gamma^{\prime}$ increases the Majorana gap~\cite{TakikawaF2020}.
In contrast, for positive $\Gamma$, the enhancement is governed by the two-spin term $\kappa_{xy}^{(M)}$.
This latter behavior cannot be straightforwardly explained by perturbation theory and may be related to possible phase transitions induced by the $\Gamma$ interaction.
By comparing the results in the quantum and classical systems, we find that the main effect of the $\Gamma^{\prime}$ interaction is well captured by the classical model, 
whereas the effect of the $\Gamma$ interaction is not.
In other words, the behavior of the three-spin term $\kappa_{xy}^{(L)}$ 
can be partly reproduced within the classical model, whereas $\kappa_{xy}^{(M)}$ cannot.
This suggests that the $\Gamma$ interaction introduces stronger quantum effects into the system compared to $\Gamma'$.

The enhanced $\kappa_{xy}/T$ with overshooting behavior is observed across a broad range of magnetic fields and interaction parameters. 
Moreover, the peak temperature of $\kappa_{xy}/T$ consistently aligns with the low-temperature peak of the specific heat, underscoring their common origin in fractional excitations. 
Our comprehensive analyses consistently support the dominant contributions of topological Majorana fermions to this overshooting behavior, rather than topological magnons. 
We note that previous studies based on the topological magnon picture predicted an opposite sign of $\kappa_{xy}/T$ for the pure Kitaev model~\cite{McClartyDGRPMP2018,Koyama2024}, although the inclusion of additional $\Gamma$ and $\Gamma'$ interactions may alter the sign~\cite{ChernZK2021,ZhangCK2021}. 
One possible origin of this discrepancy lies in the limited applicability of the magnon picture in this regime, where the magnetization remains small. 
Indeed, it was demonstrated that magnon-magnon interactions beyond the linear spin-wave theory can alter not only the amplitude but also the sign of $\kappa_{xy}/T$ in this field and temperature regime~\cite{Koyama2024}. 
Another plausible explanation involves how the edges were treated in previous calculations on topological magnons, which relied on the bulk-edge correspondence without explicitly accounting for edge-specific effects, such as modulation of the magnetic states near the edges~\cite{KoyamaN2023,HabelMWK2024}. 
Indeed, our calculations reveal that the edge magnetization differs significantly from the bulk magnetization as discussed in Appendix~\ref{app:reconstruction}.

In contrast, our calculations explicitly incorporate open boundaries, and the thermal Hall conductivity is evaluated directly from the energy current at the edge. 
More importantly, our framework does not assume any particular nature of heat-carrying quasiparticles, but instead includes all contributions within the spin model on equal footing in an unbiased manner. 
In the Kitaev model under a magnetic field, the nature of quasiparticles is expected to crossover from Majorana fermions, which are valid at zero field, to magnons, which are appropriate in the polarized regime at strong fields. 
Since this crossover remains elusive~\cite{WinterRMCHV2017,YoshitakeNKM2020}, it is essential to compute the thermal Hall response without assuming a specific quasiparticle picture. 
Our results suggest that the Majorana fermion picture is pertinent to the enhanced thermal Hall transport across a wide range of temperatures and magnetic fields, including the polarized regime beyond the quantum critical point. 
Despite ongoing debates regarding the microscopic parameters in candidate materials such as $\alpha$-RuCl$_3$, our finding provides a key insight that may help reconcile conflicting experimental observations of thermal Hall conductivity. 

Our study highlights the crucial importance of developing unbiased numerical methods to investigate thermal Hall transport in exotic states of matter in correlated electron systems. 
We demonstrate that this approach is essential, revealing that even a small magnetic field and additional interactions amounting to just a few percent of the Kitaev interaction strength can lead to behavior beyond the predictions of perturbation theory. 
This importance is particularly pronounced in systems where the nature of quasiparticles remains unknown. 
As thermal Hall measurements gain increasing relevance and are performed across a wide range of strongly correlated materials, our method based on the proper definition of the energy current offers a fundamental theoretical framework for their interpretation. 
It also serves as a powerful tool for identifying exotic quasiparticles from experimental observations. This work paves the way for a more comprehensive understanding of the thermal Hall effect in strongly correlated materials.

\begin{acknowledgments}
We wish to thank Y. Kato, K. Fukui, and K. Ido for fruitful discussions.
This work was supported by  Grant-in-Aid for Scientific Research
Nos.~JP19H05825, JP19K03742, JP20H00122, 	JP22H01179, JP22K18682, JP23K22450, JP23H03818, JP24K00563, and JP25H01247 from the Ministry of Education, Culture, Sports, Science and Technology, Japan.
It is also supported by JST CREST Grant No.~JPMJCR18T2, JST PRESTO Grant No.~JPMJPR19L5, JST COI-NEXT Program Grant No.~JPMJPF2221, and JST FOREST JPMJFR236N. This work was also supported by the National Natural Science Foundation of China (Grant No.~12150610462). Numerical calculations were performed using the facilities of the Supercomputer Center, The Institute for Solid State Physics, The University of Tokyo.
\end{acknowledgments}

\appendix
\section{Benchmark on tensor network method}
\label{app:XTRG_Bench}
\begin{figure}[t]
  \begin{center}
    \includegraphics[width=\linewidth]{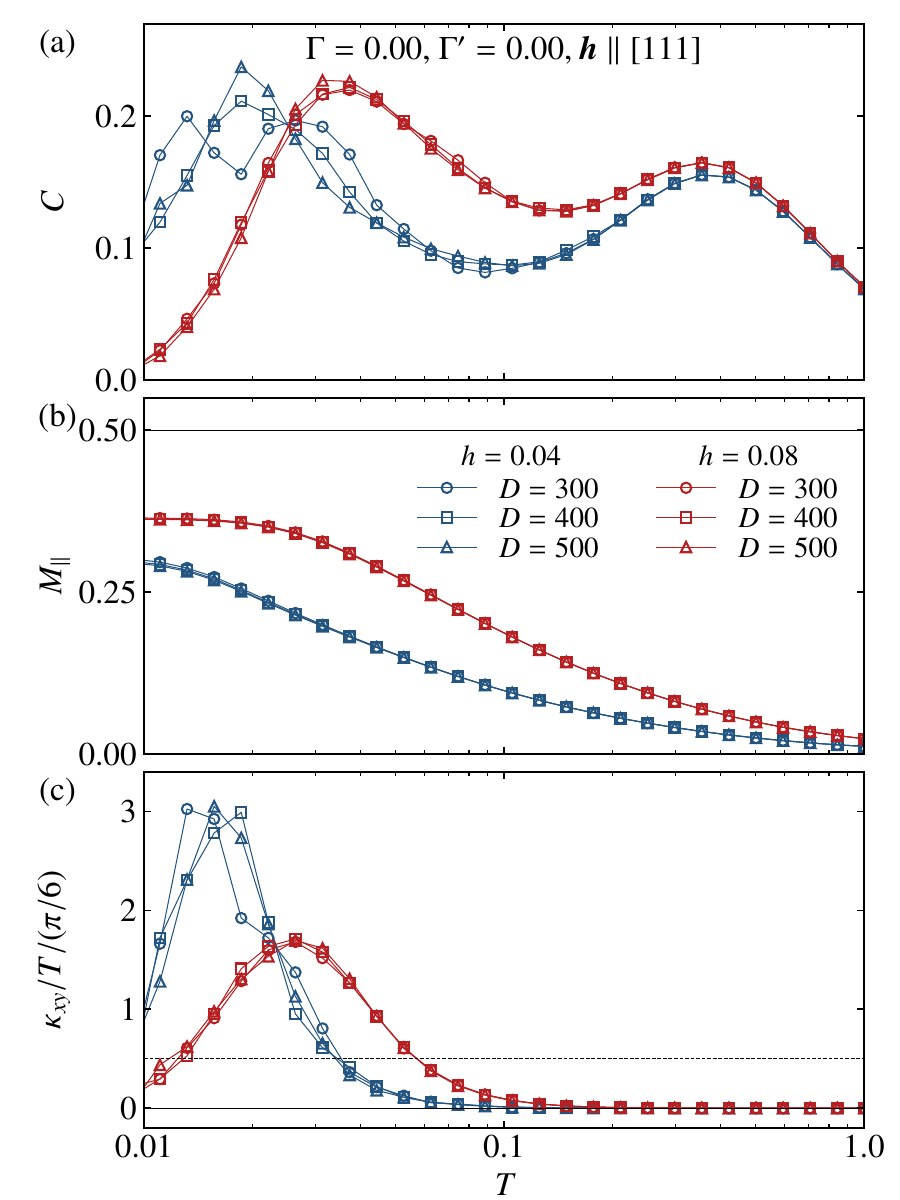}
  \end{center}
  \caption{Temperature dependence of (a) the specific heat (b) the magnetization, and (c) the thermal Hall conductivity of the pure Kitaev model under the $[111]$ magnetic field with $h=0.04$ and $h=0.08$, computed for different bond-dimensions $D=300$, $400$, and $500$. The system size is $(L, L') = (6, 6)$, consistent with that used throughout the present analysis.
}
  \label{fig:CMk_XC6}
\end{figure}

In this Appendix, we present benchmark results for the XTRG method. 
First, in Fig.~\ref{fig:CMk_XC6}, we show the dependence of representative physical quantities on the bond dimension $D$ for the pure Kitaev model under the $[111]$ magnetic field with $h=0.04$ and $0.08$ on the $(L, L') = (6, 6)$ cluster.
For $h=0.04$, while the data remain well converged across different values of $D$ down to $T\sim 0.05$, we observe noticeable deviations at lower temperatures in the specific heat and the thermal Hall conductivity. 
However, the data for $D=400$ and $500$ display consistent behaviors with slight deviations, though the data for $D=300$ exhibit distinct nonmonotonic temperature dependence in both quantities. 
From these observations, we conclude that $D=500$ is sufficiently large to capture the essential physics down to $T\sim 0.01$. 
We note that convergence with respect to $D$ improves significantly at higher fields, as demonstrated by the data for $h=0.08$ in Fig.~\ref{fig:CMk_XC6}, also supporting the use of the results with $D=500$ in the present analyses. 

\begin{figure}[t]
  \begin{center}
    \includegraphics[width=\linewidth]{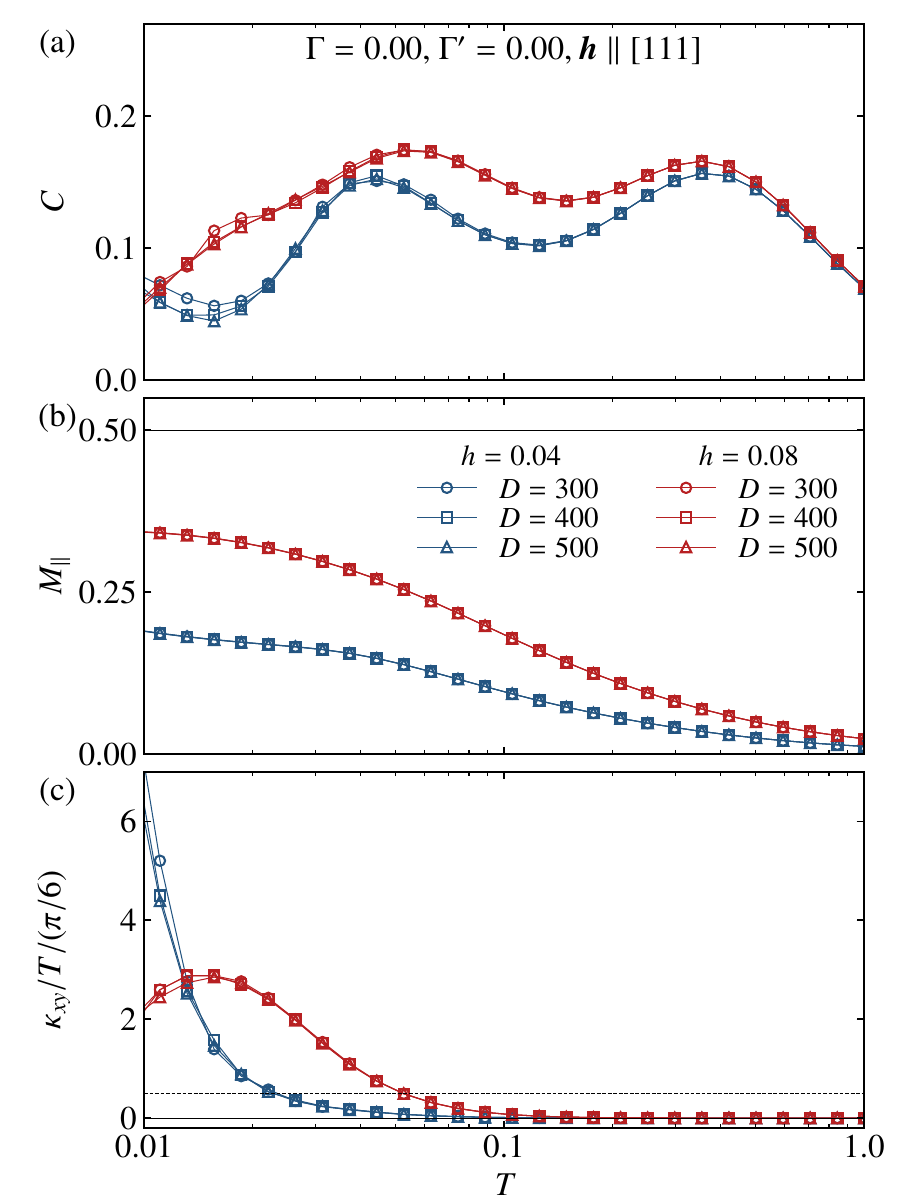}
  \end{center}
  \caption{Corresponding plot to Fig.~\ref{fig:CMk_XC6} for $(L, L') = (8, 4)$.} 
  \label{fig:CMk_XC4}
\end{figure}

Next, we examine the dependence on cluster size and geometry. Figure~\ref{fig:CMk_XC4} shows the same set of physical quantities for a different cluster with $(L, L') = (8, 4)$. Due to the smaller system size, convergence with respect to $D$ is better than the $(L, L') = (6, 6)$ case, particularly for $h=0.04$. However, the reduced system size results in a qualitatively different temperature dependence in the specific heat and the thermal conductivity; we observe an additional upturn in the specific heat around $T\simeq 0.01$ for $h=0.04$, and a significant shift of the peak of $\kappa_{xy}/T$ to lower temperatures, relative to the $(L, L') = (6, 6)$ cluster. 
The three-peak structure in the specific heat is a known artifact of small size clusters~\cite{NasuUM2015}; see also Fig.~\ref{comp_ED} in Appendix~\ref{app:cTPQ}. The absence of this additional peak in Fig.~\ref{fig:CMk_XC6}(a), which retains the well-established two-peak structure, supports the use of the $(L, L') = (6, 6)$ cluster in our main calculations. 

\section{Benchmark on thermal pure quantum state}
\label{app:cTPQ}
In this Appendix, we examine the accuracy of the cTPQ state method by comparing its results with those obtained from the ED for a small $(L, L') = (2, 4)$ cluster (the total system size is $N_{\rm s}=16$).
This serves as a preliminary step for the comparison between the cTPQ state and XTRG methods presented in Appendix~\ref{app:XTRGcTPQ}. 
In the ED approach, we diagonalize the Hamiltonian, whose dimension is
$2^{16}=65536$, using ScaLAPACK~\cite{scalapack}, and compute the temperature dependence of physical quantities based on the obtained eigenvalues and eigenvectors.
In the cTPQ state method, we employ the calculation conditions mentioned in Sec.~\ref{subsec:Thermal pure quantum state}.

\begin{figure}[t] 
\begin{center} 
\includegraphics[width=\linewidth]{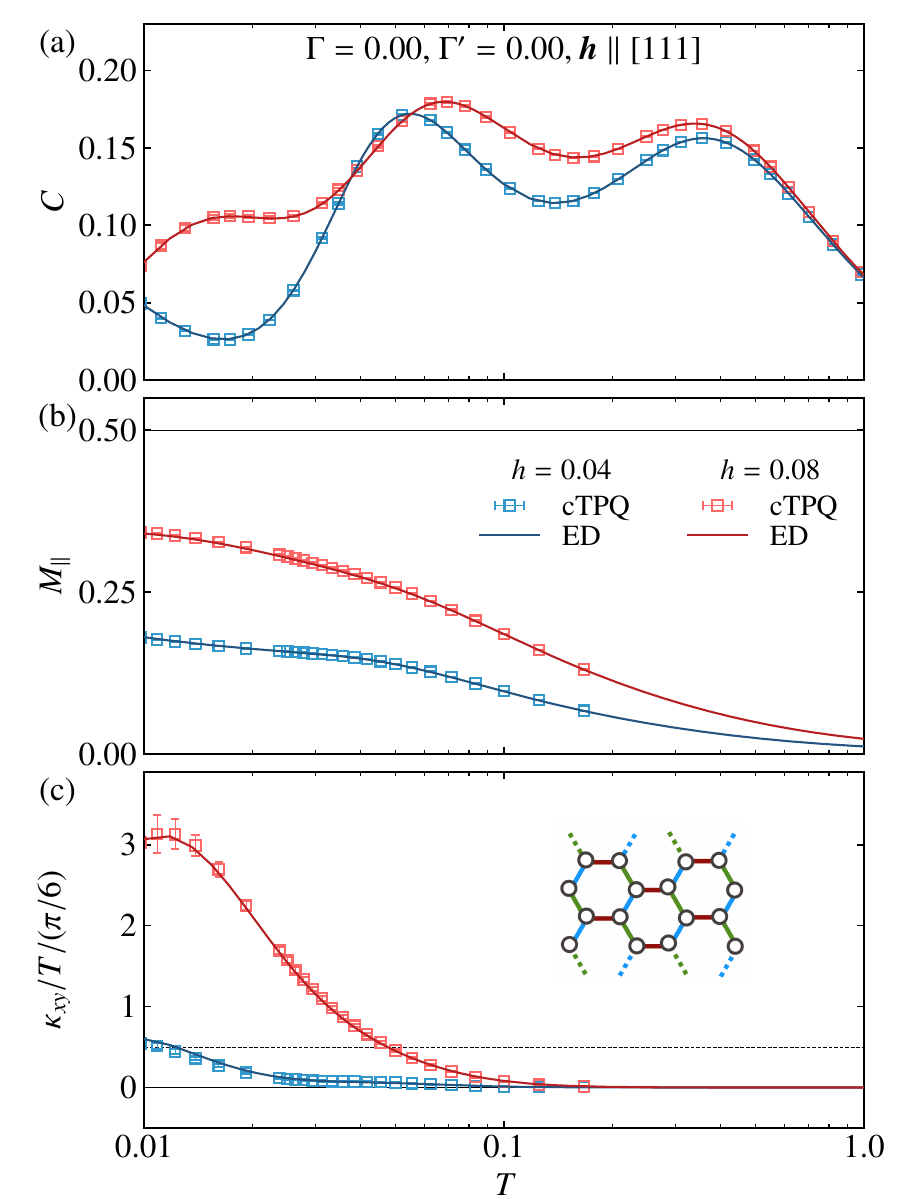}
\vspace{-0.5cm} 
\caption{Temperature dependence of (a) the specific heat, (b) the magnetization, and (c) $\kappa_{xy}/T$ of the pure Kitaev model under the $[111]$ magnetic field for $(L, L') = (2, 4)$ cluster with $h=0.04$ and $h=0.08$, calculated by the cTPQ state method and the exact diagonalization (ED) method.
In the bootstrap sampling, we take $1000$ independent initial states and choose $1000$ samples $500$ times with replacement to evaluate the average values and statistical errors. 
}
\label{comp_ED}
\end{center}
\end{figure}

The results are shown in Fig.~\ref{comp_ED}. 
For all quantities, including the thermal Hall conductivity, the cTPQ state method successfully reproduces their temperature dependencies obtained by the ED method. 
This confirms the reliability of the cTPQ approach in this temperature range. 
We note that the specific heat $C$ exhibits a three-peak structure, similar to Fig.~\ref{fig:CMk_XC4}(a). This is an artifact of small size clusters, as mentioned in Appendix~\ref{app:XTRG_Bench}. 

\section{Comparison between thermal pure quantum state method and tensor network method}
\label{app:XTRGcTPQ}
\begin{figure}[tbh] 
\begin{center} 
\includegraphics[width=\linewidth]{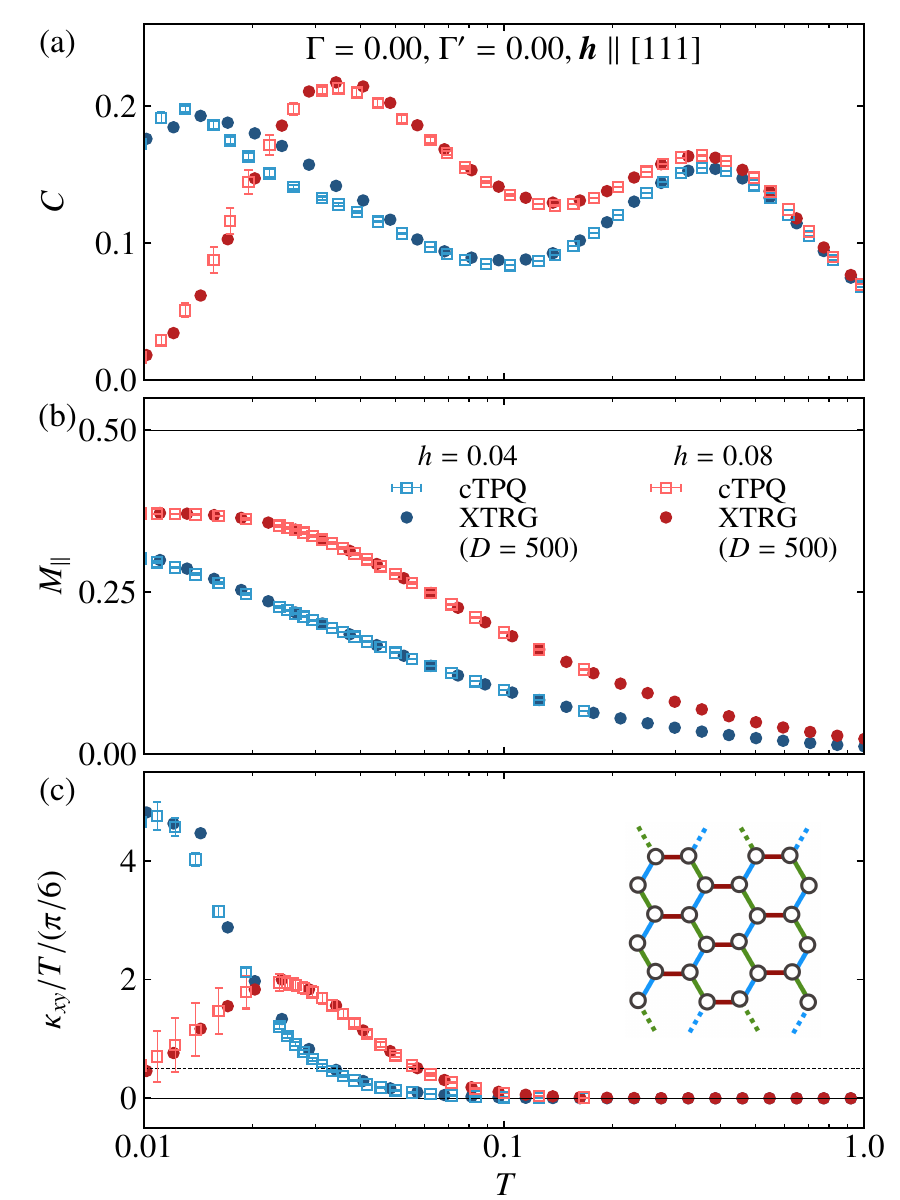}
\vspace{-0.5cm} 
\caption{Temperature dependence of (a) the specific heat, (b) the magnetization, and
(c) the thermal Hall conductivity $\kappa_{xy}$ for $(L, L') = (2, 6)$ cluster with $h=0.04$ and $h=0.08$, calcluated by the cTPQ state method and the XTRG method with $D=500$.
In the bootstrap sampling on the cTPQ results, we take $100$ independent initial states,
and choose $100$ samples $50$ times with replacement to evaluate
the average values and statistical errors. 
}
\label{comp_XTRG}
\end{center}
\end{figure}

In this Appendix, we examine the accuracy of the XTRG method by comparing its results with those obtained using the cTPQ state method for the $(L, L') = (2, 6)$ cluster ($N_{\rm s}=24$). 
In Fig.~\ref{comp_XTRG}, we show the temperature dependence of the specific heat, magnetization, and $\kappa_{xy}/T$ for $h=0.04$ and $h=0.08$, calculated using both the cTPQ state method and the XTRG method with $D=500$. 
Aside from minor discrepancies in the specific heat and $\kappa_{xy}/T$ at low temperatures for $h=0.04$, the two methods exhibit excellent agreement across two orders of magnitude in temperature, covering the range in which the main results of this study are presented. 

\section{Spatial extension of the energy current}
\label{app:J_line}
\begin{figure*}[t] 
\begin{center} 
\includegraphics[width=\linewidth]{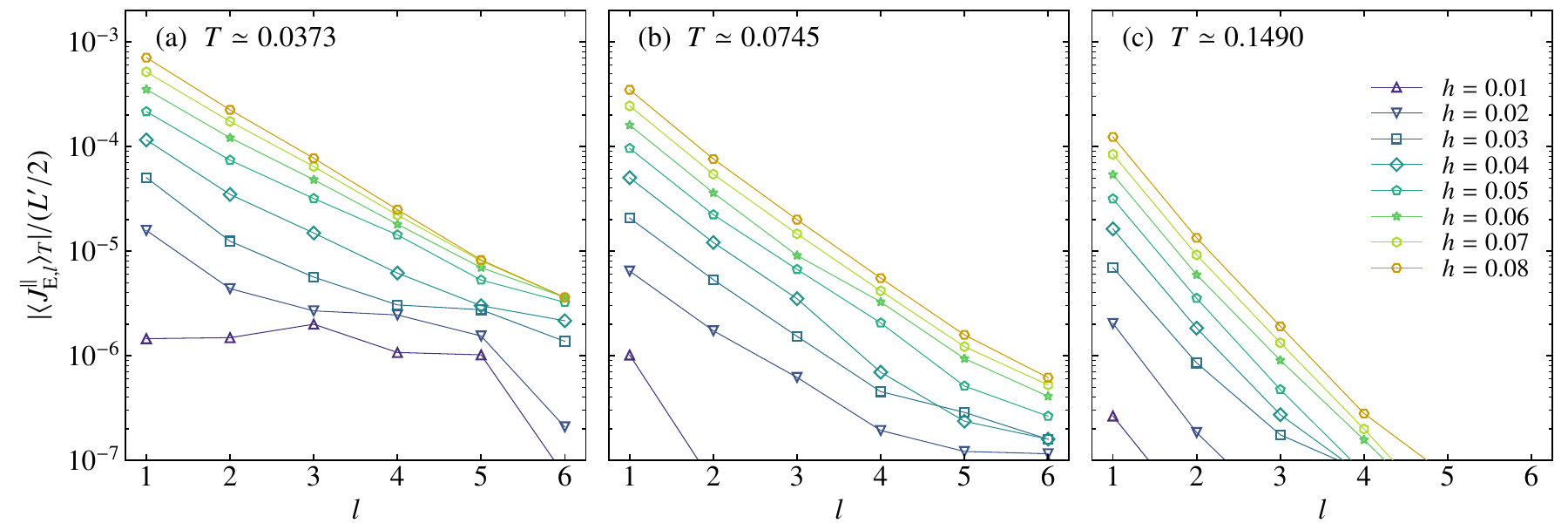}
\end{center}
\caption{Semi-log plot of the position ($l$) dependence of the energy current amplitude for the pure Kitaev model at several values of the external magnetic field parallel to the $[111]$ direction at (a) $T \simeq 0.0373$, (b) $T\simeq 0.0745$, and (c) $T\simeq 0.1490$.}
\label{fig:J_line_dep_appendix}
\end{figure*}

In this Appendix, we present the supplemental data for $l$ dependence of the energy current amplitude for the pure Kitaev model. As discussed in Sec.~\ref{subsec:Thermal Hall conductivity}, our definition of the thermal Hall conductivity $\kappa_{xy}$ relies on the assumption that the energy current is well localized at the edge. In Fig.~\ref{fig:J_line_dep}, we showed that the energy current amplitude is actually well localized at the edge for $T \simeq 0.0186$. An almost exponential decay of the energy current amplitude is also observed at higher temperatures, as shown in Fig.~\ref{fig:J_line_dep_appendix}, although the amplitude becomes smaller. These results support that the assumption underlying the definition of $\kappa_{xy}$ remains valid over a broad range of temperatures and magnetic fields.

\section{Modulation of the magnetic state near the edges}
\label{app:reconstruction}
\begin{figure}[t] 
\begin{center} 
\includegraphics[width=\linewidth]{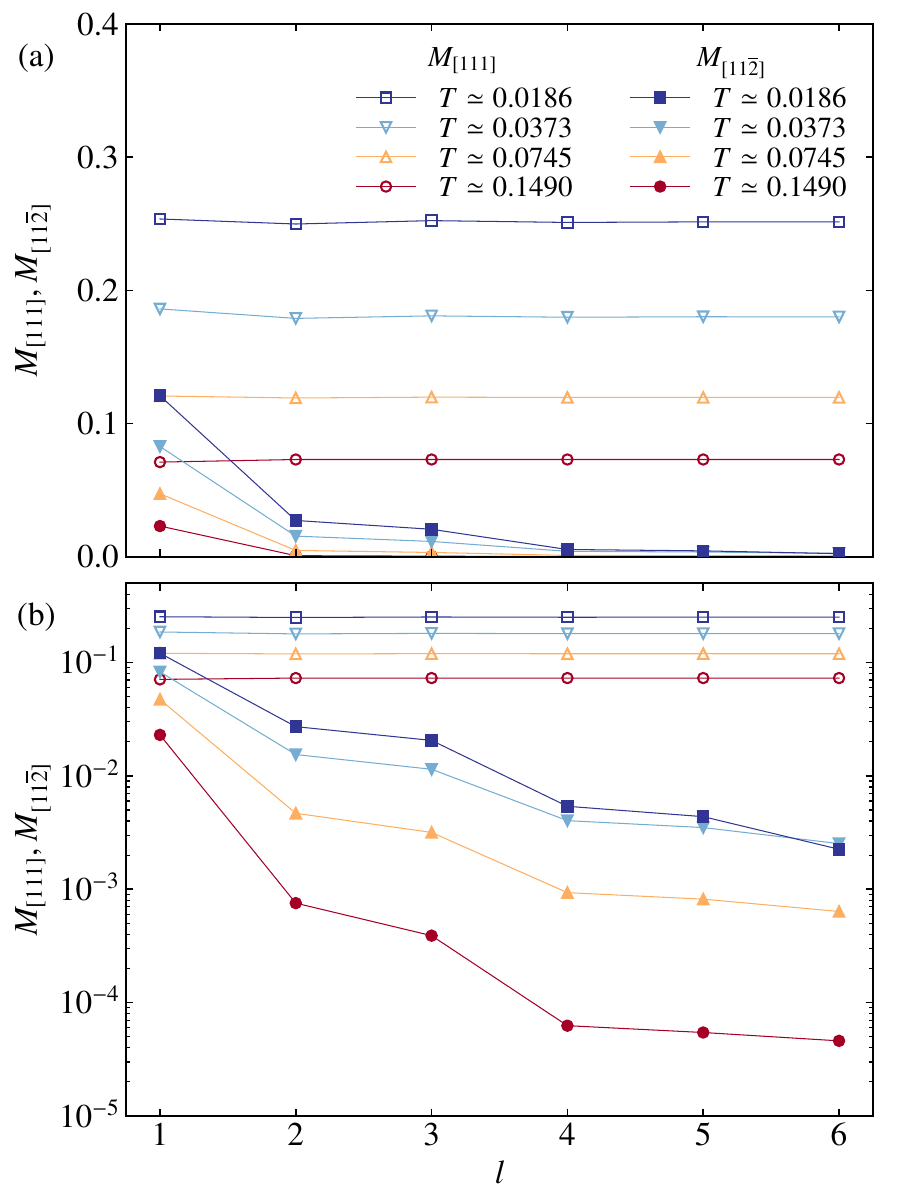}
\end{center}
\caption{The position ($l$) dependence of the local magnetization along the $[111]$ and $[11\bar{2}]$ directions for the pure Kitaev model at several values of temperature under the magnetic field parallel to the $[111]$ direction with $h=0.04$ in (a) linear scale and (b) semi-log scale. Here, the local magnetization is defined at the top sites in each position ($l$) on the lattice in Fig.~\ref{fig:lattice}(b).}
\label{fig:Edge_mag_line}
\end{figure}

In this Appendix, we discuss the magnetic state near the edges in the pure Kitaev model under the $[111]$ magnetic field. Due to the presence of open boundaries, the magnetic state near the edges can be different from the bulk magnetization. 
In Fig.~\ref{fig:Edge_mag_line}, we plot the position dependence of the local magnetization along the $[111]$ direction ($M_{[111]}$) and $[11\bar{2}]$ direction ($M_{[11\bar{2}]}$). 
Here, we define the magnetization at position $l$ as that of the top site in Fig.~\ref{fig:lattice}(b). 
In the center of the system ($l \sim 6$), $M_{[11\bar{2}]}$ almost vanishes and the magnetization is almost parallel to the $[111]$ direction, while spins near the edge $l=1$ exhibit a canting to the $[11\bar{2}]$ direction. 
This spin canting could be attributed to the absence of the $z$-bond at the edge, suggesting that it is energetically favorable to have larger $x$ and $y$ components of the magnetization. 
It is worthy noting that, similarly to the energy current amplitude, the magnetization along the $[11\bar{2}]$ direction also exhibits an almost exponential decay as a function of the position $l$ toward the center region, as shown in Fig.~\ref{fig:Edge_mag_line}(b). This indicates that this magnetic modulation may contribute to the energy current localized at the edge.

\clearpage
\bibliography{Kitaev_kxy}
\end{document}